\documentclass[nofootinbib, tightenlines, superscriptaddress,11pt]{revtex4}

\usepackage{graphicx}
\usepackage{amsmath,amssymb,amsfonts,amsthm,stmaryrd,mathtools,bm,physics}
\usepackage{color}
\usepackage{tikz}
\usepackage[normalem]{ulem}
\usepackage{braket}
\allowdisplaybreaks[1]

\def\be#1\ee{\begin{align}#1\end{align}}

\def\ba{\begin{eqnarray}}
\def\ea{\end{eqnarray}}
\def\nn{\nonumber}
\def\q{\quad}

\usepackage[bookmarks,linktocpage, colorlinks=true, plainpages = false, citecolor = treegreen,  linkcolor=darkblue, urlcolor = darkblue, filecolor = blue]{hyperref} 

\definecolor{darkblue}{rgb}{0., 0.4, 0.8}
\definecolor{treegreen}{rgb}{0., 0.7, 0.3}

\begin{document}

\title{Spikes and spines in 4D Lorentzian simplicial quantum gravity}

\author{Johanna Borissova}
\email{jborissova@perimeterinstitute.ca} 
\affiliation{Perimeter Institute, 31 Caroline Street North, Waterloo, ON, N2L 2Y5, Canada}
\affiliation{Department of Physics and  Astronomy, University of Waterloo, 200 University Avenue West, Waterloo, ON, N2L 3G1, Canada}
\author{Bianca Dittrich}
\email{bdittrich@perimeterinstitute.ca} 
\affiliation{Perimeter Institute, 31 Caroline Street North, Waterloo, ON, N2L 2Y5, Canada}
\affiliation{Theoretical Sciences Visiting Program, Okinawa Institute of Science and Technology Graduate University, Onna, 904-0495, Japan}
\author{Dongxue Qu}
\email{dqu@perimeterinstitute.ca} 
\affiliation{Perimeter Institute, 31 Caroline Street North, Waterloo, ON, N2L 2Y5, Canada}
\author{Marc Schiffer}
\email{mschiffer@perimeterinstitute.ca} 
\affiliation{Perimeter Institute, 31 Caroline Street North, Waterloo, ON, N2L 2Y5, Canada}

\begin{abstract}
Simplicial approaches to quantum gravity such as quantum Regge calculus and spin foams include configurations where bulk edges can become arbitrarily large while the boundary edges are kept small. Spikes and spines are prime examples for such configurations. They pose a significant challenge for a desired continuum limit, for which the average lengths of edges ought to become very small.
Here we investigate spike and spine configurations in four-dimensional Lorentzian quantum Regge calculus. We find that the expectation values of arbitrary powers of the bulk length are finite. To that end, we explore new types of asymptotic regimes for the Regge amplitudes, in which some of the edges are much larger  than the remaining ones. The amplitudes simplify considerably in such asymptotic regimes and the geometric interpretation of the resulting expressions involves a dimensional reduction, which might have applications to holography.

\end{abstract}

\maketitle
\tableofcontents

\section{Introduction}\label{Sec:Introduction}

The path integral approach to quantum gravity requires to construct an integral over a suitable space of geometries. Simplicial  approaches, such as quantum Regge calculus~\cite{Williams:1986hx,Hamber:2009zz} and spin foams~\cite{Perez:2012wv}, utilize triangulations as regulators. The geometry of these triangulations is specified by geometric data, e.g., edge lengths in the case of Regge calculus. The path integral is then replaced by sums over these geometric data, thus turning an a priori infinite-dimensional path integral to a finite-dimensional one. 

One could therefore compare such approaches to lattice regularizations of quantum field theories, such as lattice QCD. However, whereas in lattice QCD the lattices feature an explicit lattice constant, there is a priori no such lattice constant available in Regge gravity or spin foams. The path integral rather sums over all geometric data and thus also over the lengths of the edges in the triangulation. One could attempt to define a lattice constant by considering the expectation values of the lengths of all edges and by averaging over these expectation values. 

\begin{figure}[b]
 	\centering
 	\includegraphics[width=.7\textwidth]{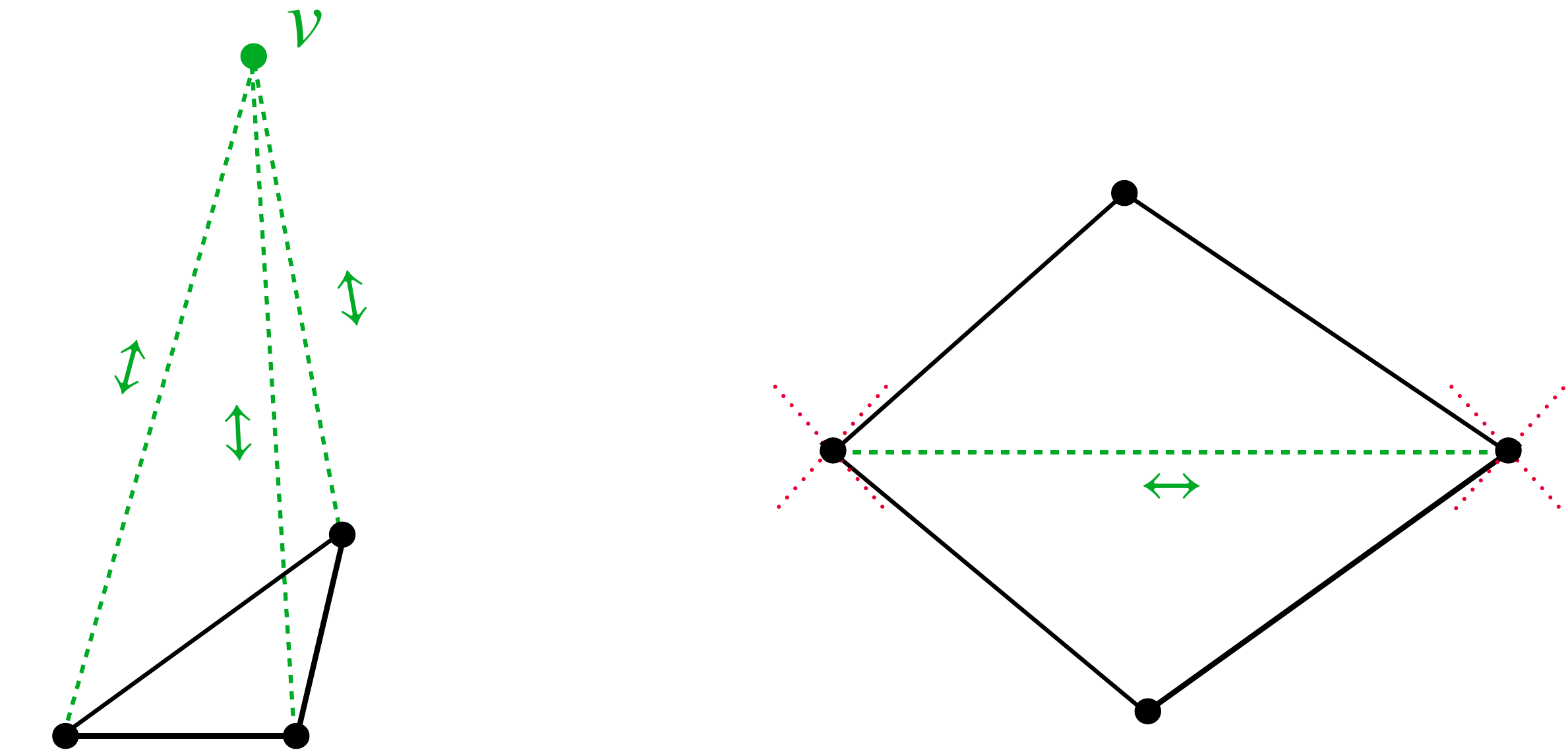}
 	\hfill
 	\caption{\label{Fig:SpikeSpine} 
 		Left: Illustration of a spike configuration in $d=2$ Lorentzian spacetime. The black solid edges form the boundary complex, while green dashed lines indicate bulk edges.  The lengths of the bulk edges can become arbitrarily large, while the lengths of the boundary edges are kept fixed. Right: Illustration of a spine configuration in $d=2$ Lorentzian spacetime. The black solid edges form the boundary complex, while the bulk edge is indicated by the dashed green line. Its  length can become arbitrarily large by tilting the boundary edges closer and closer towards the light rays marked by the dotted red lines.}
 \end{figure}

There is however a priori no guarantee that these expectation values are finite. This is due to configurations in which bulk edges can become arbitrarily large, while the boundary geometry is fixed. The prime examples of such a behaviour are so-called spikes and spines~\cite{Borissova:2024pfq}. To define a spike configuration, consider a bulk vertex $v$ and the set of all simplices containing this vertex, i.e., the star of $v$, see Figure~\ref{Fig:SpikeSpine} for an illustration. We fix the lengths of the edges in the boundary of this set to some finite values. Spikes are configurations where the length of the edges, which share the vertex $v$, can become arbitrarily large.~\footnote{The lengths of the edges are restricted by the generalized triangle inequalities, see Section~\ref{sec:gentriang}. For spikes, the generalized triangle inequalities allow infinitely large edge lengths.} 

Spine configurations can only appear in  Lorentzian triangulations. Here we consider a bulk edge $e$ and the set of all simplices containing this edge, i.e., the star of $e$, see Figure~\ref{Fig:SpikeSpine} for an illustration. We again fix the length of the edges in the boundary of this set to some finite values. A spine configuration allows for an arbitrarily large (timelike or spacelike) length of the edge $e$. 

Thus any subregion of the triangulation containing a bulk vertex or a bulk edge can include a spike or spine configuration, respectively. Spikes and spines can therefore easily dominate the path integral, and in this way preclude a suitable continuum limit. 
 
Hence, obtaining control over spike and spine configurations should be viewed as a key task in simplicial approaches to quantum gravity, such as Regge calculus and spin foams. 

In the Euclidean version of quantum Regge calculus, spike configurations are particularly problematic. It has been shown that in the two-dimensional theory~\footnote{The Regge action is a topological invariant in two dimensions. Spikes in two dimensions are often defined as configurations with arbitrarily large edge lengths, but with finite area. This can be obtained by scaling the boundary edges to be small.  The requirement for a finite area prevents an exponential suppression by the cosmological constant term. In this paper, we consider the four-dimensional theory (without a cosmological constant) and drop the requirement of a finite four-volume for the spikes.} spikes lead to divergences for the expectation values of the edge lengths~\cite{Ambjorn:1997ub}.  In three and four dimensions, spike configurations include the conformal factor mode~\cite{Dittrich:2011vz}, which makes the Euclidean action unbounded from below~\cite{Gibbons:1978ac}. The $\exp(-S_E)$ amplitude of Euclidean quantum gravity therefore leads to an exponential enhancement of spike configurations. 

Lorentzian quantum gravity could in principle avoid the conformal factor problem due to the oscillatory nature of the amplitudes. There are, however, only few investigations of this issue in Lorentzian Regge calculus, in particular for the four-dimensional case: \cite{Borissova:2023izx} shows that the conformal mode problem is indeed avoided perturbatively to one-loop order. \cite{Tate:2011ct} and~\cite{Ito:2022ycc} consider the two-dimensional theory and, inspired by Causal Dynamical Triangulations~\cite{Loll:2019rdj},  implement a restriction on the spacetime signatures of the edges in the lattice, which does not allow for spike configurations (but does allow for other configurations with large bulk edges). Furthermore, \cite{Mikovic:2023tgg} employs a measure that exponentially suppresses configurations with large edge lengths. 

Here, we will consider four-dimensional Lorentzian Regge calculus and show that a particular class of spike and spine configurations, appearing in the $5-1$ and $4-2$ Pachner moves~\cite{Pachner:1991pm}, leads to a finite path integral and to finite expectation values for arbitrary powers of the edge lengths in combination with a large class of measures (without explicit exponential suppression factors). The techniques employed here are very similar to the ones utilized for the three-dimensional theory studied in~\cite{Borissova:2024pfq}, but the present paper can be read independently from~\cite{Borissova:2024pfq}.

A key point of the present work is to analyze the Regge action in regimes where some edges are much larger than the remaining edges. We will find that the Regge action, which in general is a highly non-polynomial function of its edge lengths, simplifies considerably in the asymptotic regime. In fact, we will often find that the leading-order coefficients in the Regge action refer to the geometry of a lower-dimensional subtriangulation. We hope therefore that the techniques employed here and this phenomenon can be generalized to other configurations with large edge lengths. 

Spikes are closely related to bubbles, which are of great concern for spin foams~\cite{Perini:2008pd,Bonzom:2013ofa,Riello:2013bzw,Banburski:2014cwa,Chen:2016aag,Dona:2022vyh,Dona:2023myv} and group field theories~\cite{BenGeloun:2011jnm,Finocchiaro:2020fhl,Carrozza:2024gnh}. Spin foams differ from the Regge approach in that they are based on a different set of geometric data, namely areas and angles, which feature a discrete spectrum (more details are provided in Section~\ref{GentoN}). The strategy developed here can nevertheless help to investigate systematically bubble configurations in spin foams.

Our results on the finiteness of spike and spine configurations on the one hand provide further evidence that the Lorentzian (Regge or spin foam) path integral to quantum gravity can be made well-defined. On the other hand, we will also identify examples of spike configurations which are light-cone irregular, i.e., configurations which do not have a regular light-cone structure everywhere. This includes configurations describing topology change in time. For a certain type of such configurations, the Regge action features a branch cut and imaginary terms. A complete specification of the Lorentzian path integral therefore inevitably requires a prescription of how to deal with such configurations~\cite{deBoer:2022zka}.   

This paper is structured as follows:  In Subsection~\ref{Sec:CRegge} we introduce the complex Regge action and the notion of light cone (ir-)regular configurations. Subsection~\ref{sec:gentriang} discusses the generalized triangle inequalities. In Subsection~\ref{VolumeAsmp} we analyse the asymptotic regimes for the volumes of (sub-) simplices in the limit of one or multiple large edges. The results obtained therein will help us to derive asymptotic approximations to the Regge action for spine and spike configurations arising in the $4-2$ and $5-1$ Pachner moves, respectively, in  Section~\ref{Sec:ReggeActionAsymptotics}. In Section~\ref{Sec:PathIntegralAsymptotics}, we consider expectation values of powers of the length variables and establish their convergence properties. We close with a discussion in Section~\ref{discussion}.

\section{Lorentzian geometry of simplices}

\subsection{The complex Regge action}\label{Sec:CRegge}
 
In this section, we provide a general outline of the Lorentzian Regge action~\cite{Sorkin:1975ah,Sorkin:2019llw} and in particular employ the framework of the complex Regge action~\cite{Asante:2021phx}. A detailed overview of the geometry of Lorentzian and Euclidean simplices can be found in the Appendix of~\cite{Asante:2021phx,Borissova:2023izx}.

Regge calculus~\cite{Regge:1961px} is a discrete approach to study the  dynamics of general relativity. The $d$-dimensional spacetime manifold is approximated by gluing piecewise flat~\footnote{A framework using homogeneously curved simplices is also available~\cite{Bahr:2009qd}.}  $d$-dimensional simplices along shared $(d-1)$-dimensional subsimplices. In Lorentzian Regge calculus, the configuration variables of the Regge action are the signed squared lengths  $s_e$ for the edges $e$ of the triangulation. (Our convention for the signature of the spacetime metric is $(-,+,+,\cdots)$.) The complex Regge action is given by~\cite{Asante:2021phx}
\be\label{eq:ReggeAction}
S = - \imath  \sum_{h}\sqrt{\mathbb{V}_h}\,\epsilon_h\,,
\ee
where the sum is taken over both bulk and boundary hinges $h$, i.e., the $(d-2)$-dimensional subsimplices in the triangulation. The signed squared volume $\mathbb{V}_h$ of a hinge $h$ can be calculated using a Cayley-Menger determinant, see Section~\ref{sec:gentriang}. The bulk and boundary deficit angles, which provide a measure for the curvature concentrated at a given hinge, are defined as~\footnote{Here we make one of the two possible sign choices for the definition of the complex Regge action, see~\cite{Asante:2021phx}. Both choices lead to the same complex action, see~\cite{Asante:2021phx}. \label{footnoteA}}
 \be\label{eq:DeficitAngles}
 \epsilon_{h}^{\text{(bulk)}} = 2\pi + \sum_{\sigma \supset h} \theta_{\sigma, h}\,,\quad \epsilon_{h}^{\text{(bdry)}} = \pi   + \sum_{\sigma\supset h}\theta_{\sigma, h} \,,
 \ee
where $\theta_{\sigma,h}$ denotes the complex dihedral angle at the hinge $h$ in the simplex $\sigma$. The choice of the constant $\pi$ in the definition of the boundary deficit angles is a convention, which we will adopt throughout this paper. 

The complex dihedral angles $\theta_{\sigma,h}$ in a simplex $\sigma$ at a hinge $h\subset \sigma$ are defined by~\cite{Asante:2021phx,Jia:2021xeh}
\ba\label{eq:DihedralAngleLog}
\theta_{\sigma,h} = -\imath \log \qty(  
\frac{ \vec{a}\cdot \vec{b} -\imath \sqrt{ (\vec{a}\cdot \vec{a}) (\vec{b}\cdot \vec{b})- (\vec{a}\cdot \vec{b})^2   }
}
{ \sqrt{\vec{a}\cdot \vec{a}}\sqrt{\vec{b}\cdot \vec{b}}}
)\,,
\ea
with
\ba\label{equ4}
\vec{a}\cdot \vec{b}=
\frac{d^2}{  \mathbb{V}_{h}}  \frac{\partial \mathbb{V}_{\sigma}}{\partial s_{\bar{h}}} \, ,\q\q
\vec{a}\cdot \vec{a}=  \frac{\mathbb{V}_{\rho_a} }{   \mathbb{V}_{h}  }  \, ,\q\q
\vec{b}\cdot \vec{b}=  \frac{\mathbb{V}_{\rho_b} }{   \mathbb{V}_{h}  } \, .
\ea
Here, $\rho_a \subset \sigma$ and $\rho_b \subset \sigma$ are the $(d-1)$-subsimplices of $\sigma$ sharing the hinge $h$, and $\bar{h} \subset \sigma$ is the edge opposite to $h$. We shall adopt the 
principal branches for square roots and logarithms, however, we also need to specify~\footnote{This specification is determined by the choice of convention mentioned in the previous footnote.} which sides of the branch cuts for square roots and logarithms to use. For the remainder of this article, along the negative real axis, we define $\sqrt{-r} = \imath \sqrt{r}$ and $\log(-r) = \log(r) - \imath \pi$ where $r \in \mathbb{R}_+$. Thus we adopt the branch-cut value originating from the lower complex half-plane for the logarithm and from the upper complex half-plane for the square root.

The complex dihedral angles allow to represent Euclidean as well as Lorentzian angles. In a Lorentzian triangulation, both Euclidean or Lorentzian dihedral angles can occur. More precisely, if the hinge $h$ is spacelike, the space orthogonal to $h$ includes a timelike direction and the associated dihedral angle is Lorentzian. Conversely, if $h$ is timelike, the associated dihedral angle is Euclidean. Null hinges do not contribute to the Regge action (\ref{eq:ReggeAction}), as their volume is zero.

If the data $\{\vec{a}\cdot\vec{b},\vec{a}\cdot\vec{a},\vec{a}\}$ defines a Euclidean angle, the complex angle in (\ref{eq:DihedralAngleLog}) and the Euclidean angle are related by $\theta_{\sigma,h}=-\psi^E_{\sigma,h}$. As a consequence, the (bulk) deficit angle for a timelike hinge $h$ is $\epsilon^E_h=2\pi-\sum_{\sigma \supset h}\psi^E_{\sigma,h}$. As the volume square $\mathbb{V}_h$ for a (timelike) hinge is negative, the contribution of timelike hinges to the Regge action (\ref{eq:ReggeAction}) is real. 

For a Euclidean triangulation which satisfies the Euclidean generalized triangle inequalities (see Section~\ref{sec:gentriang}), we can also define dihedral angles (\ref{eq:DihedralAngleLog}). The Euclidean Regge action (which provides a discretization of the Euclidean action for general relativity) is then given by
\begin{align}
S^E = -\sum_{h}\sqrt{\mathbb{V}_h}\epsilon^E_h\,.
\end{align}
In this case the complex Regge action in (\ref{eq:DihedralAngleLog}) evaluates to $S=\imath S^{E}$. Thus, the complex Regge action $S$ for a Euclidean signature triangulation is purely imaginary.

\begin{figure}[t]
	\centering
	\includegraphics[width=.9\textwidth]{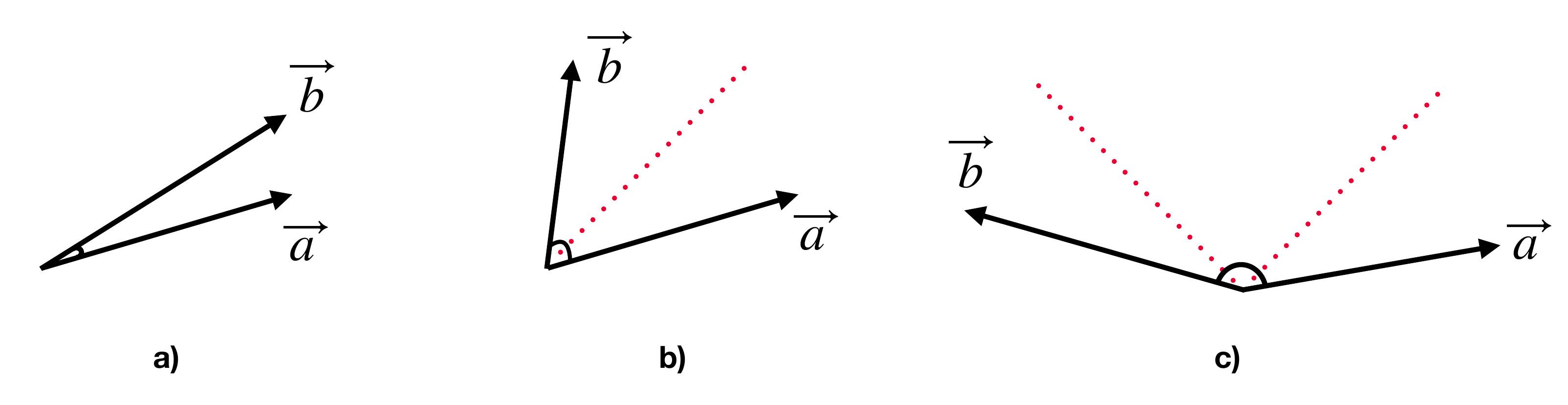}
	\hfill
	\caption{\label{Fig:angleab} 
The Lorentzian angle between two vectors $\vec{a}$ and $\vec{b}$. The dotted red lines represent the enclosed light rays. The value of the Lorentzian angle between $\vec{a}$ and $\vec{b}$ is $\left(\beta_{\sigma,h}-   \imath m_{\sigma,h} \,\pi/2\right)$, where $\beta_{\sigma,h} \in \mathbb{R}$ and $m_{\sigma,h}\in\{0,1,2\}$ depends on the number of light rays included in the (convex) wedge between $\vec{a}$ and $\vec{b}$. }
\end{figure}  

Let us return to the Lorentzian triangulation and consider a spacelike hinge. The data $\{\vec{a}\cdot\vec{b},\vec{a}\cdot\vec{a},\vec{a}\}$ then defines a Lorentzian angle, i.e., $\vec{a}$ and $\vec{b}$ can be embedded into two-dimensional Minkowski space. According to~\cite{Sorkin:2019llw,Asante:2021phx}, the complex angles in (\ref{eq:DihedralAngleLog}) can be expressed in terms of Lorentzian angles as $\theta_{\sigma,h}=-\imath \psi^{L}_{\sigma,h}=-\imath \left(\beta_{\sigma,h} -\imath m_{\sigma,h} \pi/2\right)$, where $\psi^{L}_{\sigma,h}$ is the Lorentzian angle, $\beta_{\sigma,h} \in \mathbb{R}$, and $m_{\sigma,h}\in \{0,1,2\}$ indicates the number of light rays included in the (convex) wedge between $\vec{a}$ and $\vec{b}$, as illustrated in Figure~\ref{Fig:angleab}. Therefore, the Lorentzian (bulk) deficit angle is given by:
\ba
\epsilon^L_{\sigma,h}= 2\pi-\frac{\pi}{2}\left(\sum_{\sigma \supset h} m_{\sigma_h} \right) \,-\,\imath \sum_{\sigma \supset h} \beta_{\sigma,h} \, . 
\ea
This angle is purely imaginary if $\sum_{\sigma \supset h} m_{\sigma_h} = 4$, i.e., if the sum of the dihedral angles includes exactly four light rays and therefore two light cones. We will refer to spacelike hinges satisfying this condition as \emph{light-cone regular}. Timelike hinges are light-cone regular by definition. Therefore, the contribution of a regular hinge to the Regge action (\ref{eq:ReggeAction}) is real. However, if the number of light ray crossings in the bulk deficit angle associated with a given hinge $h$, denoted as
\ba
{\cal N}_h=\sum_{\sigma \supset h} m_{\sigma_h} \,,
\ea
is smaller or larger than $4$, we will obtain a negative or positive imaginary contribution to the Regge action, respectively. 

It is important to note that the sign of the imaginary contributions for the Lorentzian Regge action depends on the choice of convention, mentioned in Footnote~\ref{footnoteA}. This choice is equivalent to defining the Lorentzian action by using a complexification of the squared edge lengths   $s_e \rightarrow s_e +\imath \varepsilon$ (so that one avoids the branch cuts of the square roots and the logarithm) and taking the limit $\varepsilon \rightarrow 0$~\cite{Asante:2021phx}. Using $s_e \rightarrow s_e -\imath \varepsilon$ instead, we obtain, in the case that all hinges are light-cone regular, the same value for the Lorentzian action as with $s_e \rightarrow s_e +\imath \varepsilon$. However, in the case of an irregular hinge, the sign of the imaginary term of the action changes. This shows that the complex Regge action has branch cuts if the data define a Lorentzian triangulation and if there are light-cone irregular hinges. The Regge action on opposite sides of the branch cut associated to a given hinge, differs only in the sign of the imaginary term originating from this hinge. Thus, if we perform a path integral, we can choose the sign in front of these imaginary terms, as this amounts to a choice along which side of the branch cut the path integral contour is placed~\cite{Asante:2021phx,Dittrich:2024awu}.

Note that the imaginary terms are determined by the analytical structure of the Regge action and can be reproduced by analytical continuation  from two different starting points:   The first one is to apply a generalized Wick rotation and construct the Lorentzian action through an analytic continuation from the Euclidean action~\cite{Asante:2021phx}. An alternative starting point is to begin with Lorentzian data which are light-cone regular, and analytically continue to data with light-cone irregularities by slightly deviating into complexified squared edge lengths to go around branch points, as discussed in~\cite{Asante:2021phx}. 

Light-cone irregular hinges represent a particular type of conical singularities for a Lorentzian metric, that leads to imaginary terms for the action also in the continuum~\cite{Louko:1995jw, Neiman:2024znb}. Such conical singularities play an important role for the derivation of thermodynamic quantities such as entropies, from the gravitational path integral~\cite{Marolf:2022ybi, Dittrich:2024awu}. For two-dimensional spacetimes, such conical singularities seem to appear always for spacetimes describing topology change~\cite{Sorkin:2019llw}. In three- and four-dimensional triangulations, light-cone irregular hinges can also appear without topology change~\cite{Dittrich:2021gww}.

\subsection{Generalized triangle inequalities} \label{sec:gentriang}

The generalized Euclidean and Lorentzian triangle inequalities ensure that a simplex with given signed length squares can be embedded into flat Euclidean or Minkowski space, respectively. The generalized triangle inequalities for a Euclidean or spacelike (non-degenerate) simplex $\sigma$ require that the signed squared volume for the simplex $\sigma$ itself and all its sub-simplices $\rho$ satisfy the following condition:
\ba 
\mathbb{V}_{\sigma} >0,  \quad \mathbb{V}_{\rho}>0\,,
\ea 
where the signed squared volume of a $d$-dimensional simplex $\sigma^d=(012\cdots d)$ is given by:
\ba
\mathbb{V}_{\sigma^d}&\,=\,& \frac{ (-1)^{d+1} }{ 2^{d} (d!)^2} \det  \left(
\begin{matrix}
	0 & 1 & 1 & 1 & \cdots & 1 \\
	1 & 0 & s_{01} & s_{02} & \cdots & s_{0d} \\
	1 & s_{01} & 0 & s_{12} & \cdots & s_{1d} \\
	\vdots & \vdots & \vdots & \vdots & \ddots & \vdots \\
	1 & s_{0d} & s_{1d} & s_{2d} & \cdots & 0
\end{matrix}
\right)\, ,
\ea
where $s_{ij}$ is the signed squared length of the edge between vertices $i$ and $j$. Moreover, the signed squared volume determines the spacelike, null, or timelike nature of a simplex. A simplex $\rho$ is timelike if $\mathbb{V}_{\rho} < 0$, null if $\mathbb{V}_{\rho} = 0$, and spacelike if $\mathbb{V}_{\rho} > 0$.

The generalized triangle inequalities for a Lorentzian $d$-simplex $\sigma^d$ in $d$-dimensional spacetime~\eqref{eq:Inequalities} require that if a subsimplex $\rho\subset \sigma^d$ is timelike or null, then all subsimplices $\rho' \subset \sigma^d$ containing this subsimplex $\rho$, i.e., $\rho\subset \rho'$, must not be spacelike~\cite{Tate:2011rm}. That is, a timelike simplex cannot be a subsimplex of a spacelike simplex. Therefore, for a Lorentzian $d$-simplex $\sigma$, we have the condition that $\mathbb{V}_{\sigma^d} < 0$ (non-degeneracy) and additionally that: 
\be\label{eq:Inequalities}
\rho \subset \sigma \,,\,\mathbb{V}_\rho \leq 0 \quad \Rightarrow \quad \forall \rho' \supset \rho: \,\mathbb{V}_{\rho'}\leq 0\,.
\ee

We will consider the asymptotics of the Regge action with one or multiple bulk edges scaled to become large. The generalized triangle inequalities determine whether certain edges are allowed to become large. E.g., for a Euclidean $d$-simplex, we cannot scale just one of the edges to be large while keeping the other edges fixed, as this violates the Euclidean triangle inequality. For a Lorentzian $d$-simplex, such a scaling is however possible. For example, let us consider a timelike triangle $(012)$ with a spacelike (or timelike) edge $(01)$ and with the other edges timelike (or spacelike). The squared area of triangle $(012)$ is given by
\ba
\mathbb{V}_{(012)}=
-\frac{1}{16} \left[s_{01}^2+   (s_{02}-s_{12})^2-2s_{01}\left(s_{02}+s_{12}\right)\right]<0 \, ,
\ea 
which shows that the Lorentzian triangle inequality is always satisfied. Therefore, we can scale the edge $(01)$ to become large while still satisfying the triangle inequality.

\subsection{Simplex geometry in the limit of large edges}\label{VolumeAsmp}

We will consider  triangulations with a few simplices and with boundary, in the limit where the bulk edges are large as compared to the boundary edges. We will thus have simplices involving one or more edges which are much larger than the remaining edges. 

To compute the Regge action in this limit, we need to
 investigate the asymptotic behavior of the dihedral angles. To this end, we will first derive the asymptotic behavior of the volumes for a $d$-simplex and its subsimplices, as the volume and the dihedral angles satisfy the relations~\cite{Dittrich:2007wm}
\ba \label{SinCos} 
\sin(\theta_{\sigma, h}) \,\,=\,\, -\frac{d}{d-1}\frac{\sqrt{\mathbb{V}_h} \sqrt{\mathbb{V}_\sigma}}{\sqrt{\mathbb{V}_{\rho_a}}\sqrt{\mathbb{V}_{\rho_b}}}  \,,\q\q\q
\cos(\theta_{\sigma, h}) \,\,=\,\, \frac{d^2 }{\sqrt{\mathbb{V}_{\rho_a}}\sqrt{\mathbb{V}_{\rho_b}}} \pdv{\mathbb{V}_\sigma}{s_{\bar{h}}} \, ,
\ea
where we use the same notation as for equation~(\ref{equ4}).
These relations can be derived from (\ref{eq:DihedralAngleLog}), see ~\cite{Borissova:2023izx}. Moreover, the derivative of the squared volume with respect to a squared edge length can be expressed in terms of the volume squares of the simplex and its subsimplices~\cite{Dittrich:2007wm,Borissova:2023izx}:
\ba
\left(\frac{\partial \mathbb{V}_\sigma}{\partial s_{i j}}\right)^2=\frac{1}{d^4} \mathbb{V}_{\bar{i}} \mathbb{V}_{\bar{j}}-\frac{1}{d^2(d-1)^2} \mathbb{V}_\sigma \mathbb{V}_{\overline{i j}}\, .
\ea
Here we denote with $\sigma$ a $d$-simplex $(0,\ldots,(d-1))$, $s_{ij}$ is the squared lengths of the edge $(ij)$,  $\mathbb{V}_{\bar{i}}$ is the squared volume of the subsimplex of $\sigma$, which we obtain by removing the vertex $i$ from $\sigma$, and $\mathbb{V}_{\bar{ij}}$ is the squared volume of the subsimpplex of $\sigma$, which we obtain by removing the vertices $i$ and $j$ from $\sigma$.

Thus, to derive the asymptotic behavior of the dihedral angles we essentially need the asymptotic behavior of the volumes of the simplex and its subsimplices.

To study the asymptotic behavior of the volumes of a $d$-dimensional simplex $\sigma^d = (0\cdots d)$ with vertices $\{0,\dots,d\}$, we will use the expression for the signed squared volume via the determinant of the associated Cayley-Menger matrix $C$ in terms of the signed squared edge lengths~\cite{Sorkin:1975ah}:
\ba\label{eq::VolumeCaleyMenger}
\mathbb{V}_{\sigma^d}&\,=\,& -\frac{ (-1)^{d} }{2^{d} (d!)^2} \det  \left(
\begin{matrix}
	0 & 1 & 1 & 1 & \cdots & 1 \\
	1 & 0 & s_{01} & s_{02} & \cdots & s_{0d} \\
	1 & s_{01} & 0 & s_{12} & \cdots & s_{1d} \\
	\vdots & \vdots & \vdots & \vdots & \ddots & \vdots \\
	1 & s_{0d} & s_{1d} & s_{2d} & \cdots & 0
\end{matrix}
\right) \equiv -\frac{ (-1)^{d} }{ 2^{d} (d!)^2} \det(C)  \,.
\ea
Applying Laplace's expansion for the determinant of a $(d+2)\times (d+2)$ matrix $C$ and expanding around an arbitrary row $i$, we can write
\be\label{eq:LaplaceDet}
\det(C) = \sum_{j=1}^{d+2} (-1)^{i+j}C_{ij} \det(\tilde{C}_{ij})\,.
\ee
Here, $C_{ij}$ is the $(ij)$-th entry of $C$, and $\tilde{C}_{ij}$ denotes the determinant of the submatrix of $C$ obtained by removing the $i$-th row and $j$-th column of $C$. With the expansion (\ref{eq:LaplaceDet}), it is straightforward to separate the terms that include large edge lengths and to determine the asymptotic behavior.

\subsubsection{Volumes in the limit of one large edge}

Here, we will study the asymptotic behavior of the squared volume $\mathbb{V}_{\rho^d}$ of a $d$-simplex $\rho^d=(012\cdots d)$ with only one large edge $(01)$. The squared volume $\mathbb{V}_{\rho^d}$ is a polynomial of at most quadratic order in $s_{01}$:
\be
\mathbb{V}_{\rho^d} = a s_{01}^2 + b s_{01} + c\,,
\ee
where $a$, $b$ and $c$ do not depend on $s_{01}$. By using Laplace's formula~\eqref{eq:LaplaceDet} repeatedly, we can derive the first coefficient as
\be\label{eq:VolumeScalingD-2Moves}
\mathbb{V}_{(012\cdots d)} = -\frac{1}{4 d^2(d-1)^2} \mathbb{V}_{\overline{01}}s_{01}^2 + {\cal O}(s_{01}^{1})\,,
\ee
where $\mathbb{V}_{\overline{01}}$ is the signed squared volume of the subsimplex of $\rho^d$ obtained by removing the vertices $(0)$ and $(1)$. Thus, $\mathbb{V}_{\rho^d}$ is of order $s_{01}^2$ if $\mathbb{V}_{\overline{01}} \neq 0$. From (\ref{eq:VolumeScalingD-2Moves}), one can see that the signature of the squared $d$-dimensional volume $\mathbb{V}_{\rho^d}$ depends on the signature of $\mathbb{V}_{\overline{01}}$. If the $d$-simplex $(012\cdots d)$ is timelike (spacelike), the $(d-2)$-simplex $(23\ldots d)$ needs to be spacelike (timelike) to satisfy the triangle inequalities when the edge $(01)$ is large. Furthermore, one can also see that scaling only one edge length to be large is only possible for timelike (or null) simplices. 

Formula (\ref{eq:VolumeScalingD-2Moves}) involves a dimensional reduction: the leading-order coefficient of the volume of the $d$-simplex $(012\cdot d)$ is determined by the volume of its $(d-2)$-subsimplex $(2\cdot d)$. We will see later, that for 4-simplices, this extends to the dihedral angles at triangles $(01k)$ (with $k=2,3,4$) which contain the large edge --- the leading-order coefficient of a given dihedral angle will coincide with the angle in the triangle $(234)$ at the vertex $(k)$.

\subsubsection{Volumes in the limit of multiple large edges} \label{VolumeAsmpMulti}

Next we will study the asymptotic behavior of the squared volume $\mathbb{V}_{\rho^d}$ of a $d$-simplex $\rho^d=(012\cdots d)$, which has $d$ large edges $(0i)$, where $i\in \{1,\ldots,d\}$. There are two options for scaling multiple edges to become large: one can either choose a multiplicative scaling $s_{0i}=\lambda s_{0i}^0$, or an additive scaling $s_{0i}=s_{0i}^0 \pm \lambda$ and then consider the limit $\lambda \rightarrow \infty$. Here we choose the additive scaling throughout, so that the leading-order term of the volumes does not depend on the initial values $s_{0i}^0$. This will lead to simpler results for the asymptotic form of the action.  With the additive scaling, also the triangle inequalities become independent of the initial values, whereas, for the multiplicative scaling, the triangle inequalities can lead to intricate conditions on the $s_{0i}^0$~\cite{Borissova:2024pfq}.

We start by considering a triangle $(012)$, whose signed squared area can be written as
\ba\label{eq:V012expansion}
\mathbb{V}_{(012)} &=&
 -\frac{1}{16}s_{01}^2 - \frac{1}{16}s_{02}^2 + \frac{1}{8}s_{01}s_{02} + \frac{1}{8}s_{01}s_{12}+ \frac{1}{8}s_{02}s_{12}-\frac{1}{16}s_{12}^2\,.
\ea

In the case that the two large edges have the same signature, $s_{01}=s_{02}=\pm\lambda$, the terms quadratic in $\lambda$ cancel out, and we are left with
\ba\label{Vol2dA}
\mathbb{V}_{(012)} &=&\pm\frac{1}{4}\lambda s_{12} + \mathcal{O}(\lambda^0)\, . 
\ea

If we consider $s_{01}=\pm\lambda$ and $s_{02}=\mp \lambda$, the dominant term in~\eqref{eq:V012expansion} is quadratic in $\lambda$:
\ba\label{Vol2dB}
\mathbb{V}_{(012)} &=& -\frac{1}{4}\lambda^2  + \mathcal{O}(\lambda^1) \, . 
\ea

 In general, for a $d$-simplex in which the large edges $s_{0i}=\pm \lambda$ ($i=1,\ldots,d)$ agree in their signature, we obtain
 \ba\label{Volhom}
\mathbb{V}_{(012\cdots d)} &=&\pm\frac{1}{d^2}\mathbb{V}_{(12\cdots d)} \lambda  + \mathcal{O}(\lambda^0)\, . 
 \ea
 From this we see that, if all large edges are timelike, the $d$-simplex must be timelike (or null), and the subsimplex  $(12\cdots d)$ must be spacelike (or null). In the case where all large edges are spacelike, the signature of the $d$-simplex must agree with the signature of the subsimplex $(12\cdots d)$. Note that for a four-dimensional Lorentzian triangulation (with non-degenerate top-dimensional simplices) all 4-simplices must be timelike.

 We again notice that (\ref{Volhom}) involves a dimensional reduction, this time, from $d$-simplices $(01\cdot d)$ to $(d-1)$-subsimplices $(1\cdot d)$. 
 This will again extend to the dihedral angles. E.g., for a 4-simplex $(01234)$ we have that the leading-order coefficient of the dihedral angle at a triangle  $(0ij)$ will be given by the dihedral angle at the edge $(ij)$ in the tetrahedron $(1234)$.

In the case that $s_{01}=\pm \lambda$ and $s_{0j}=\mp\lambda$ for $j=2,\ldots, d$, we find
\ba\label{Volnonhom1}
\mathbb{V}_{(012\cdots d)} &=& -\frac{1}{d^2\times (d-1)^2} \mathbb{V}_{(23\cdots d)}\lambda^2 + \mathcal{O}(\lambda^1) \, .
\ea
Clearly, the $d$-simplex $(012\cdots d)$ has to be timelike, and therefore the subsimplex $(23\cdots d)$ has to be spacelike. 

We again notice a dimensional reduction in (\ref{Volnonhom1}) from a $d$-simplex to a $(d-2)$-simplex. For a 4-simplex $(01234)$ this reduction extends only to the dihedral angles at triangles $(01i)$ with $i=2,3,4$. It does not apply to dihedral angles at $(0kl)$ with $2\leq k<l\leq 4$, as these triangles contain two large edges of the same signature.

Allowing for a renaming of the vertices, the cases described in (\ref{Volhom}) and (\ref{Volnonhom1}) cover all possible choices for the signatures of the edges $(0i)$ in a triangle $(012)$ and in a tetrahedron $(0123)$. 

For a 4-simplex, there is, however, in addition also the case $s_{01}=s_{02}=\pm \lambda$ and $s_{03}=s_{04}=\mp\lambda$. In this case one can verify by an explicit calculation that the following relations hold:
\ba\label{Vol4dC}
\mathbb{V}_{(01234)}
&=&\frac{\lambda^2}{2^4 \cdot (4\cdot 3)^2}\bigg(
(s_{13}-s_{14}-s_{23}+s_{24})^2 -4 s_{12}s_{34}
\bigg)+ \mathcal{O}(\lambda^1)\nn\\
&=& -\frac{\lambda^2}{2^4} \left( \frac{\partial}{\partial s_{13}}  +   \frac{\partial}{\partial s_{14}}  +  \frac{\partial}{\partial s_{23}}+\frac{\partial}{\partial s_{24}}     \right)\mathbb{V}_{(1234)}   \nn\\
&=&
\frac{ \lambda^2}{2^4}\left(2 \frac{\partial}{\partial s_{12}}\mathbb{V}_{(1234)} -\frac{1}{3^2}\mathbb{V}_{(134)} -\frac{1}{3^2}\mathbb{V}_{(234)}  \right)+ \mathcal{O}(\lambda^1) 
\nn\\
&=&\frac{ \lambda^2}{2^4}\left(2\frac{\partial}{\partial s_{34}}\mathbb{V}_{(1234)} -\frac{1}{3^2}\mathbb{V}_{(123)}- \frac{1}{3^2}\mathbb{V}_{(124)}  \right)+ \mathcal{O}(\lambda^1)  \, .
\ea
As $\mathbb{V}_{(01234)}$ has to be negative, the inequality $4 s_{12}s_{34}>(s_{13}-s_{14}-s_{23}+s_{24})^2$ must be satisfied in order to be able to apply this scaling. In particular, the signatures of $s_{12}$ and $s_{34}$ have to agree.

\section{Regge action asymptotics}\label{Sec:ReggeActionAsymptotics}

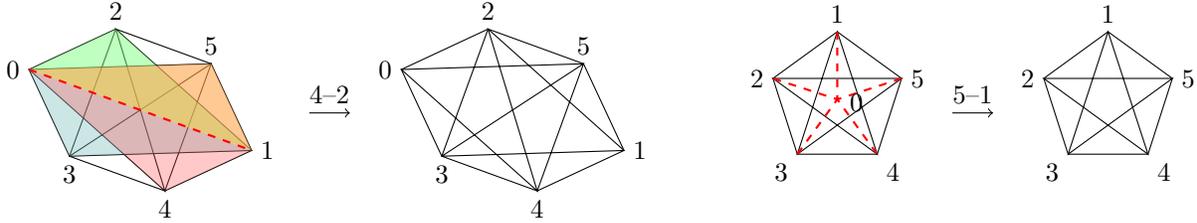
\begin{figure}[t]
	\begin{tikzpicture}[scale = 0.9]
		\begin{scope}[rotate = -20]
			\draw[] (0,1)--(-1,0)--(0,-1)--(1.5,-1)--(2.5,0)--(1.5,1)--cycle;
			\draw[]   (1.5,1)--(-1,0)--(1.5,-1)--(0,1)--(2.5,0)--(0,-1)--(0,1) (1.5,-1)--(1.5,1)--(0,-1);
			\path[fill= green!50,opacity= 0.5] (-1,0)--(2.5,0)--(0,1);
			\path[fill= teal!50,opacity= 0.4] (-1,0)--(2.5,0)--(0,-1);
			\path[fill= orange!70,opacity= 0.6] (-1,0)--(2.5,0)--(1.5,1);
			\path[fill=red!70,opacity= 0.3] (-1,0)--(2.5,0)--(1.5,-1);
			\draw[thick,red,dashed] (-1,0)--(2.5,0);    
			\node [left] at (-1,0) {$0$};
			\node [above] at (0,1) {$2$};
			\node [below] at (0,-1) {$3$};
			\node [below] at (1.5,-1) {$4$};
			\node [above] at (1.5,1) {$5$};
			\node [right] at (2.5,0) {$1$};
		\end{scope}
		
		\draw[->] (3.2,-.3)--(3.8,-0.3);
		
		\node[above] at (3.5,-0.3) {4--2};
		
		\begin{scope}[xshift=5.5cm,rotate = -20]
			\draw[] (0,1)--(-1,0)--(0,-1)--(1.5,-1)--(2.5,0)--(1.5,1)--cycle;
			\draw[]   (1.5,1)--(-1,0)--(1.5,-1)--(0,1)--(2.5,0)--(0,-1)--(0,1) (1.5,-1)--(1.5,1)--(0,-1);
			\node [left] at (-1,0) {$0$};
			\node [above] at (0,1) {$2$};
			\node [below] at (0,-1) {$3$};
			\node [below] at (1.5,-1) {$4$};
			\node [above] at (1.5,1) {$5$};
			\node [right] at (2.5,0) {$1$};
		\end{scope}

		\begin{scope}[xshift=11cm,yshift=-0.1cm]
			\draw[] (0,1)--(-0.95,0.31)--(-0.59,-0.81)--(0.59,-0.81)--(0.95,0.31)--cycle;
			\draw[]   (-0.95,0.31)--(0.95,0.31)--(-0.59,-0.81)--(0,1)--(0.59,-0.81)--cycle;
			\draw[dashed,red,thick]  (0,1)--(0,0)--(-0.95,0.31) (-0.59,-0.81)--(0,0)--(0.59,-0.81) (0.95,0.31)--(0,0);
			\node [right] at (0.05,-0.045) {$0$};
			\node [above] at (0,1) {$1$};
			\node [right] at (0.95,0.31) {$5$};
			\node [below right] at (0.59,-0.81) {$4$};
			\node [below left] at (-0.59,-0.81) {$3$};
			\node [left] at (-0.95,0.31) {$2$};

			\draw[->] (1.7,-0.2)--(2.3,-0.2);
			
			\node[above] at (2,-0.2) {5--1};
			
		\end{scope}
		
		\begin{scope}[xshift=15cm,yshift=-0.1cm]
			\draw[] (0,1)--(-0.95,0.31)--(-0.59,-0.81)--(0.59,-0.81)--(0.95,0.31)--cycle;
			\draw[]   (-0.95,0.31)--(0.95,0.31)--(-0.59,-0.81)--(0,1)--(0.59,-0.81)--cycle;
			\node [above] at (0,1) {$1$};
			\node [right] at (0.95,0.31) {$5$};
			\node [below right] at (0.59,-0.81) {$4$};
			\node [below left] at (-0.59,-0.81) {$3$};
			\node [left] at (-0.95,0.31) {$2$}; 
			
		\end{scope}
	\end{tikzpicture}
	\caption{\label{Fig:4DMoves}Four-dimensional Pachner moves $4-2$ and $5-1$.}
\end{figure}

We will now consider the asymptotic behaviour of the Regge action for triangulations with a boundary and with one or several bulk edges. We will consider the limit where these bulk edges are much larger than the boundary edges. 

More precisely, we will consider the initial configurations of the $4-2$ and $5-1$ Pachner moves, which contain one and five bulk edges, respectively. By scaling these bulk edges to become large, we obtain examples for spine configurations in the case of $4-2$ moves and spike configurations in the case of $5-1$ moves.

The initial configuration for the $4-2$ Pachner move is given by four 4-simplices, which altogether share one bulk edge, see Figure~\ref{Fig:4DMoves} for an illustration.  The boundary of the initial configuration can also serve also as a boundary of two 4-simplices sharing a tetrahedron (if the triangle inequalities are satisfied for these two 4-simplices). This serves as the final configuration of the $4-2$ Pachner move.

The initial configuration for the $5-1$ Pachner move is given by five 4-simplices, which share one bulk vertex, see Figure~\ref{Fig:4DMoves} for an illustration. This initial configuration includes five bulk edges. The boundary of the initial configuration can also serve as the boundary of one 4-simplex (if this 4-simplex satisfies the generalized triangle inequalities), which serves as the final configuration for the $5-1$ Pachner move.

The final configurations of these Pachner moves, i.e., two 4-simplices sharing a tetrahedron or a single 4-simplex, can be embedded into flat Minkowskian space, if the generalized Lorentzian triangle inequalities hold for the 4-simplices in the final configuration. We can then construct a classical solution for the (squared) lengths of the bulk edges in the initial configurations. One uses the embedding of the final Pachner move configuration into flat space to compute the lengths of the bulk edges in the initial configuration. As one uses a flat embedding, the deficit angles at all bulk hinges (i.e., bulk triangles) vanish. This satisfies the Regge equation of motion.  The Regge action for the initial configuration evaluated on this flat solution is then equal to the Regge action for the final configuration.

There are cases where the Lorentzian generalized triangle inequalities are satisfied for the initial configuration of a Pachner move (with some range of edge lengths allowed for the bulk edges), but not for the final configuration. It might still be possible that the triangle inequalities are satisfied for simplices with a different spacetime signature, e.g., Euclidean signature. In this case one can construct a solution to the equation of motion, which, however, describes a simplicial complex with this different signature. Such solutions can play a role in the path integral as saddle points in a complexified configuration space, see for instance~\cite{Dittrich:2021gww,Asante:2021phx,Dittrich:2023rcr}.

In Lorentzian signature there are more possibilities to scale the bulk edges large than in Euclidean signature. E.g., in Euclidean signature we cannot scale just one bulk edge large, as this would violate the Euclidean triangle inequality. In contrast, this is allowed in Lorentzian signature, and leads to spines in the initial $4-2$ Pachner move configurations.

The initial $5-1$ Pachner move configurations allow for unbounded bulk edge lengths, both in Euclidean and Lorentzian signature. In Lorentzian signature we can have cases where all bulk edges are either spacelike or timelike and large. But we can also have cases where some of the bulk edges are spacelike and the remainder is timelike. These mixed-signature cases, and also the case with only spacelike bulk edges, can lead to light cone irregular bulk triangles, where the action acquires imaginary terms.  We will comment more on the appearance of such imaginary terms in the discussion section~\ref{discussion}.

\subsection{$4-2$ Pachner move}

The initial $4-2$ Pachner move configuration includes four 4-simplices $(01ijk)$ with $2\leq i<j<k\leq 5$. These share a (bulk) edge $(01)$, cf.~Figure~\ref{Fig:4DMoves}. In a path integral we can integrate out the corresponding length squared variable and in this way remove the edge. The final configuration of this Pachner move can therefore be interpreted as two 4-simplices $(02345)$ and $(12345)$ glued along the tetrahedron $(2345)$. 
The classical solution can be constructed by embedding the glued final 4-simplices  $(02345)$ and $(12345)$ into Minkowski space and by determining the distance between the vertices $(0)$ and $(1)$. The two simplices can only be embedded into Minkowski space if the generalized Lorentzian triangle inequalities are satisfied. If this is not the case, there may be a complex solution. E.g., in case the two 4-simplices satisfy the Euclidean triangle inequalities, there will be a solution for the bulk edge length, which will lead to a triangulation where all four initial 4-simplices satisfy the Euclidean triangle inequalities.

We will consider the asymptotic regime for the $4-2$ configuration when the bulk edge length is very large and describes a spine configuration.

To determine the geometry of one of the four initial 4-simplices, consider an edge $e_{01}$ in a 4-simplex $(01234)$ whose edge lengths we scale to be large, $s_{01}\rightarrow \pm 
\infty $, in a triangle where the other two edge lengths remain bounded. This is only possible due to the Lorentzian triangle inequality and thus the triangle has to be timelike in the asymptotic regime. The dihedral angles at the triangles $(01i)$, $i=2,3,4$, are therefore Euclidean angles. 

The dihedral angles at $(01i)$ can be found by projecting the 4-simplex onto the plane perpendicular to the triangle $(01i)$. In the asymptotic limit, the resulting triangle assumes the same geometry as the triangle $(ijk)$. Indeed, using the asymptotic behaviour of simplex volumes derived in Section~\ref{VolumeAsmp}, we obtain for example for the dihedral angle $\theta_{(01234),(012)}$,
\ba
    \sin(\theta_{(01234),(012)}) &=& -\frac{4}{3} \frac{\sqrt{-\frac{1}{16}s_{01}^2}\sqrt{-\frac{1}{576}\mathbb{V}_{(234)}s_{01}^2}}{\sqrt{-\frac{1}{144}s_{23} s_{01}^2}\sqrt{-\frac{1}{144}s_{24}s_{01}^2}} \,+\,\mathcal{O}\qty(s_{01}^{-1}) \nn\\
   & =& \sin(\theta_{(234),(2)})\,+\,\mathcal{O}\qty(s_{01}^{-1}), \quad s_{01}\to \pm \infty\,,
	\ea
 and 
 \ba
  \cos(\theta_{(01234),(012)}) &=& \frac{16}{\sqrt{-\frac{1}{144}s_{23} s_{01}^2}\sqrt{-\frac{1}{144}s_{24}s_{01}^2}} \frac{-s_{01}^2}{576}\frac{\partial \mathbb{V}_{(234)}}{\partial s_{34}}\,+\,\mathcal{O}\qty(s_{01}^{-1}) \nn\\
  &=&\cos(\theta_{(234),(2)}) \,+\,\mathcal{O}\qty(s_{01}^{-1}) , \quad s_{01}\to \pm \infty\,.
 \ea
Similarly, the dihedral angle at the boundary triangles $(0ij)$ and $(1ij)$, such as $\theta_{(01234),(023)}$, approaches a constant for large $s_{01}$, which can be seen from the  following equation:
\ba
    \sin(\theta_{(01234),(023)}) &=& -\frac{4}{3} \frac{\sqrt{\mathbb{V}_{(023)}}\sqrt{-\frac{1}{576}\mathbb{V}_{(234)}s^2_{01}}}{\sqrt{-\frac{1}{144}s_{23}s^2_{01}}\sqrt{\mathbb{V}_{(0234)}}}+\mathcal{O}(s_{01}^{-1})\,.
\ea
Furthermore, the dihedral angle at the boundary triangles $(234)$, such as $\theta_{(01234),(234)}$, can be expressed as
\ba
    \sin(\theta_{(01234),(234)}) &=& -\frac{4}{3} \frac{\sqrt{\mathbb{V}_{(234)}}\sqrt{-\frac{1}{576}\mathbb{V}_{(234)}s^2_{01}}}{\sqrt{\mathbb{V}_{(0234)}}\sqrt{\mathbb{V}_{(1234)}}}+\mathcal{O}(s_{01}^{0})\,.
\ea
Thus $\theta_{(01234),(234)}$ scales as $\mathcal{O}(\log s_{01})$. In the Regge action, these boundary dihedral angles are multiplied by terms of order $\mathcal{O}(s_{01}^0)$. As a result, the leading contribution to the Regge action arises from the bulk deficit angles, which are multiplied by the area of the bulk triangles. 

In the  $4-2$ move configuration, there are four bulk triangles with asymptotically equal areas given by
\ba
\sqrt{ \mathbb{V}_{(01i)}}= \sqrt{-s^2_{01}}/4 + \mathcal{O}(s_{01}^0)\,,
\ea
with $i=2,3,4,5$.  Each bulk triangle $(01i)$ is shared by three 4-simplices $(01ijk)$, and the dihedral angle $\theta_{(01ijk)(01i)}$ approaches the value of the dihedral angle $\theta_{(ijk)(i)}$ in the triangle $(ijk)$. The total sum of dihedral angles $\theta_{(01ijk)(01i)}$ is thus asymptotically equal to the sum of all dihedral angles in the four triangles $(ijk)$ of the tetrahedron $(2345)$, which amounts to $-4\times\pi$. Therefore, we have
\ba
 S^{4-2} &=&-\imath \frac{ \sqrt{-s^2_{01}}}{4}(4\times 2\pi - 4\times \pi)  + \mathcal{O}(\log s_{01}) \,=\,   \pi  \abs{s_{01}} + \mathcal{O}(\log s_{01}) \, . \label{Regge42}
\ea
 Different from the $3-2$ Pachner move in three dimensions, discussed in~\cite{Borissova:2024pfq}, here we do not have an asymptotic regime with light-cone irregular bulk triangles. The reason for this is that, as discussed above, the bulk triangles are necessarily timelike in the asymptotic regime. Thus, the deficit angles at these triangles are Euclidean and \emph{cannot} be light-cone irregular.

\begin{figure}[t]
	\centering
	\includegraphics[width=.48\textwidth]{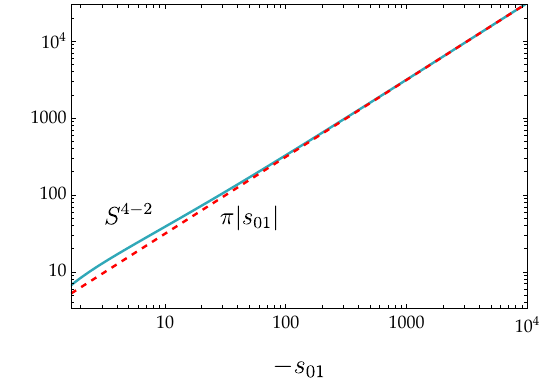}
	\hfill
	\includegraphics[width=.49\textwidth]{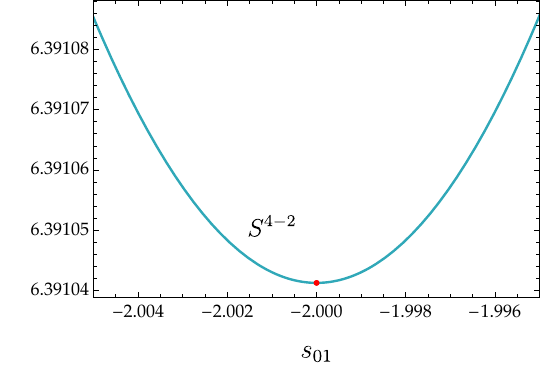}
	\caption{\label{Fig:42timelikebulk}  
Here, we consider four identical 4-simplices sharing a timelike bulk edge whose squared edge length asymptotes to infinity. We choose all $s_{ij}=2$ (for $2\leq i\leq j\leq 5$) and $s_{0i}=s_{1j}=1/4$ (for $2\leq i\leq 5$). In the left panel, we show a logarithmic plot of the exact Regge action $S^{4-2}$ (cyan curve) and its asymptotic approximation, see (\ref{Regge42}), (red dashed curve) as a function of $s_{01}\in[-10^{4}, -\frac{5}{3}]$. Note that the imaginary part of $S^{4-2}$ vanishes. In the right panel, we show the exact Regge action $S^{4-2}$ (cyan curve) as a function of $s_{01}$ in the small regime $s_{01}\in[-1.995, -2.005]$ to show that the timelike case allows for a flat solution at $s_{01}=-2$ (red point).}
\end{figure}

\begin{figure}[t]
	\centering
	\includegraphics[width=.48\textwidth]{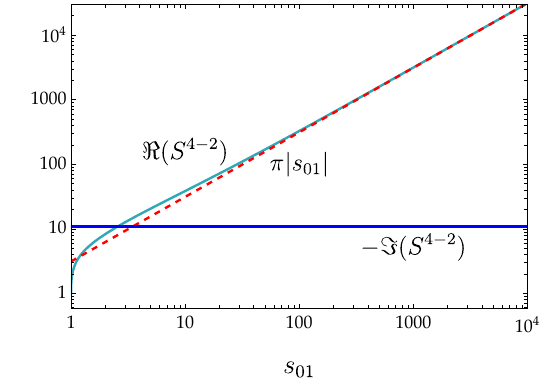}
	\hfill
	\includegraphics[width=.49\textwidth]{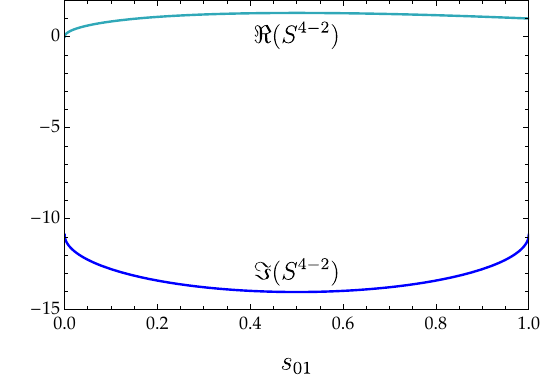}
	\caption{\label{Fig:42spacelikebulk} Here, we consider four 4-simplices sharing a spacelike bulk edge $s_{01}>0$. We choose all $s_{ij}=2$ (for $2\leq i\leq j\leq 5$) and $s_{0i}=s_{1j}=1/4$ (for $2\leq i\leq 5$). In the left panel, we show a logarithmic plot of the real part of the exact Regge action $S^{4-2}$ (cyan curve) and its asymptotic approximation, see (\ref{Regge42}), (red dashed curve) as a function of $s_{01}\in[1,10^4]$, which is the light-cone regular regime where the imaginary part of $S^{4-2}$ (blue curve) remains constant. In the right panel, we show the real and imaginary parts of the exact Regge action $S^{4-2}$ (cyan and blue curves) as a function of $s_{01}$ in the regime $s_{01}\in(0,1]$. It is evident that $s_{01}\in(0,1]$ describes light-cone irregular configurations.}
\end{figure}

~\\
{\it Example:} 
\begin{itemize}
\item  Consider an initial $4-2$ Pachner move configurations with simplices $(01ijk)$, where $2\leq i<j<k\leq 5$. We choose all $s_{ij}=2$ (for $2\leq i<j\leq 5$)  and $s_{0i}=s_{1i}=1/4$ (for $2\leq i\leq 5$). (We take all squared lengths to be in Planck units.) The generalized Lorentzian triangle inequalities restrict the squared length of the bulk edge $(01)$ to be either $s_{01}<-5/3$ or to be spacelike, $s_{01}>0$. The timelike case allows for a flat solution at $s_{01}=-2$. The spacelike case includes a light-cone irregular regime for $0<s_{01}\leq 1$. For both, the timelike and spacelike asymptotic regime, we find $S^{4-1}(s_{01})/|s_{01}|=\pi$ in the limit $|s_{01}|\rightarrow \infty$. Figures~\ref{Fig:42timelikebulk} and~\ref{Fig:42spacelikebulk} illustrate the behaviour of the Regge action for the cases with timelike and spacelike bulk edge, respectively. 
\end{itemize}

\subsection{Generalization to $N$ 4-simplices sharing an edge}\label{GentoN}

The result for the asymptotic limit of the Regge action for the $4-2$ move can be easily generalized~\footnote{We thank Jos\'e Padua-Arg\"uelles for sharing this result.} to configurations of $N$ simplices, which share an edge $e_{01}$, whose squared length we take to $\pm \infty$.  Each bulk triangle leads to a contribution of $\imath \,2\pi |s_{01}|/4$ to the action. The asymptotic values for the dihedral angles at the bulk triangles in a given 4-simplex add up to $-\pi$. These are then multiplied by the square root of the asymptotic area value $\imath |s_{01}/4$. 
We thus have
\ba\label{NSimpl}
 S &=&   \pi  \abs{s_{01}}  \frac{(2T - N)}{4} + \mathcal{O}(\log s_{01}) \,
, 
\ea
where $N$ is the number of 4-simplices which share the edge $(01)$, and $T$ is the number of (bulk) triangles, which share the edge $(01)$.

This asymptotic behaviour of the Regge action may have interesting consequences for spin foams.
As discussed in the introduction, spin foams~\cite{Perez:2012wv} are another path-integral approach based on a triangulation to which one assigns geometric data, which are then summed over in the path integral. In the case of spin foams, the geometric data are triangle areas and 3D dihedral angles~\cite{Dittrich:2008va}, which can be encoded into an area metric associated to each 4-simplex~\cite{Dittrich:2023ava}. The 3D dihedral angles can be integrated out, leaving only the areas~\cite{Asante:2020qpa}. Using triangle areas as basic variables leads to a configuration space, which is genuinely larger than the configuration space of length Regge calculus. The Regge action can however be extended to such an area configuration space~\cite{Barrett:1997tx,Asante:2018wqy}, and reduces to the length Regge action in case the area configuration is induced by a length configuration~\cite{Asante:2021zzh}. 

Spin foams suppress  configurations outside the length configuration space. This is most evident in the construction of the effective spin foam models~\cite{Asante:2020qpa,Asante:2020iwm,Asante:2021zzh}. One can thus parameterize spin foam variables by length parameters associated to the edges of the triangulation, and another set of variables describing deviations from the length configuration space, see e.g.~\cite{Dittrich:2021kzs,Dittrich:2022yoo,Han:2021kll}. The length parameters are allowed to become large, but larger deviations from the length configuration space lead to an exponential suppression of the amplitude. We will therefore restrict to configurations where these deviations are very small.

The regime in which all areas are large $A\gg1$ (we express all geometric quantities in Planck units ), allows for a semi-classical approximation. In this approximation, the 4-simplex amplitude reproduces the cosine of the Regge action~\cite{Kaminski:2017eew,Liu:2018gfc,Simao:2021qno}. (Effective spin foams make more direct use of the Area Regge action, thus the amplitude reproduces the Regge action also for smaller areas.) Here, we will consider a configuration of $N$ 4-simplices sharing a bulk edge, and a regime where all edges are large $l_e\gg1$, with the bulk edge much larger than the boundary edges. We can thus apply the asymptotic formula (\ref{NSimpl}).

We furthermore note that the areas for timelike triangles in spin foams have a discrete spectrum~\footnote{Areas are conjugated to dihedral angles~\cite{Dittrich:2008ar, Dittrich:2014rha}, which for timelike triangles are Euclidean, and therefore periodic. This explains a discrete equidistant spectrum.}, the areas have absolute values in $\mathbb{N}/2 \ell_P^2$~\cite{Conrady:2010kc}. 

With the asymptotic approximation of the absolute value of the bulk areas given by $A\sim |s_{01}|/4$ we can express~\eqref{NSimpl} as $S\sim A\pi(2T-N)$. In the case of the $4-2$ move we have $2T-N=4$. We would therefore expect that the amplitude is well approximated by $1$.  Here we ignored measure terms, which might suppress the amplitude for large values of the area, see the discussion in~\cite{Dittrich:2011vz,Dittrich:2014rha}.

\subsection{$5-1$ Pachner move}\label{Sec:51move}

In the initial $5-1$ Pachner move configuration, there are five $4$-simplices $(0ijkl)$ with $i<j<k<l$ taking values in $\{1,2,3,4,5\}$. These share a (bulk) vertex $(0)$, cf.~Figure.~\ref{Fig:4DMoves}. In a path integral, the five bulk edges with edge squared variables $s_{0i}$, $i=1,2,3,4,5$ are integrated out, and in this way the bulk vertex is removed. The remaining edges are associated with a triangulation consisting of one $4$-simplex $(12345)$. To derive the asymptotic behavior of the dihedral angles and the Regge action for the $5-1$ move, we have to distinguish between all possible combinations of signatures for the five bulk edges. 

The equations of motion for the five bulk edge lengths allow for a flat configuration. If  the $4$-simplex $(12345)$ satisfies the generalized Lorentzian~\footnote{If the 4-simplex satisfies the generalized Euclidean triangle inequalities, one can construct an Euclidean solution, which amounts to a saddle point in the complexified configuration space, see~\cite{Dittrich:2021gww} for an example.} triangle inequalities, it can be embedded into Minkowski space. Embedding also a vertex $(0)$ and adopting as lengths for the bulk edges $(0i)$ the geodesic distance between $(i)$ and $(0)$, leads to a four-parameter family of solutions. This family represents a gauge orbit with flat configurations, arising from a remnant diffeomorphism symmetry~\cite{Dittrich:2008pw,Bahr:2009ku}. The Regge action is constant along this gauge orbit and coincides with the Regge action of the 4-simplex $(12345)$. Away from the flat solution, we identify the level sets of the Regge action, which are generically four-dimensional, as gauge orbits. Thus, the path integral associated to the $5-1$ move can be simplified to a one-dimensional integral.
 
We will consider a gauge fixing $|s_{0i}|=\lambda$ for $i=1,\ldots,4$ for the asymptotic regime.  This choice corresponds to the additive scaling discussed in Section~\ref{VolumeAsmpMulti}. With this choice we assume that the gauge conditions $|s_{0i}|=\lambda$ define a good gauge fixing for large $\lambda$. 

We will proceed by considering the case of homogeneous signature for all the bulk edges.

~\\ 	\noindent
{\bf Case $s_{0i}=\pm \lambda$ (with the same sign for all $i=1,\ldots,5$):}\\\noindent
To determine the dihedral angles at the bulk edges, let us first consider the dihedral angle at the triangle $(012)$ in the 4-simplex $(01234)$:
\ba
	\sin(\theta_{(01234),(012)}) \,=\, -\frac{4}{3} \frac{\sqrt{\mathbb{V}_{012}}\sqrt{\mathbb{V}_{01234}}}{\sqrt{\mathbb{V}_{0123}}\sqrt{\mathbb{V}_{0124}}}
 &=&-\frac{3}{2}
 \frac{ \sqrt{\pm s_{12}\lambda }\sqrt{\pm \mathbb{V}_{1234}\lambda}}{\sqrt{\pm\mathbb{V}_{123}\lambda}\sqrt{\pm\mathbb{V}_{124}\lambda}} 
 +\mathcal{O}(\lambda^{-1}) \nn\\
 &=& \sin(\theta_{(1234),(12)})+\mathcal{O}(\lambda^{-1})\, .
	\ea
Likewise, we obtain $\cos(	\theta_{(01234),(012)})= \cos(	\theta_{(1234),(12)})+\mathcal{O}(\lambda^{-1})$. Similarly, the dihedral angles at the boundary triangles $(ijk)$, such as $\theta_{(01234),(123)}$, become constant for large $\lambda$. In the Regge action, these boundary dihedral angles are multiplied by terms of order $\mathcal{O}(\lambda^0)$. As a result, the dominant contribution to the Regge action comes from the bulk deficit angles, which are multiplied by the areas of the bulk triangles.

For the deficit angles at the bulk triangles $(0ij)$ we find
\ba \label{eq:epsilon0ij51}
	\epsilon_{(0ij)}^{5-1} &  =&
 2\pi + \theta_{(0ijkl),(0ij)} +   \theta_{(0ijkm),(0ij)}+ \theta_{(0ijlm),(0ij)}\nn \\
	&=& 2\pi + \theta_{(ijkl),(ij)} +\theta_{(ijkm),(ij)} +\theta_{(ijlm),(ij)}\, +\,\mathcal{O}(\lambda^{-1}) \, .
	\ea
To compute the Regge action we multiply these deficit angles with 
\ba
\sqrt{\mathbb{V}_{0ij}}=\sqrt{\pm \lambda s_{ij}/4} +\mathcal{O}(\lambda^0) \, .
\ea

We therefore obtain for the Regge action
\ba\label{eq:51moveasymptotic1}
S^{5-1} &=&
-\frac{\imath}{2}\sqrt{\lambda}\!\!\sum_{0<i<j\leq 5} \sqrt{\pm s_{ij}} \,\epsilon_{(ij)}\, \,+\,\,\mathcal{O}(\lambda^0) \, ,
\ea
where we denote with $\epsilon_{(ij)}$ the 3D deficit angle 
\ba \label{eq:3Ddeficit}
\epsilon_{(ij)}&=&2\pi + \theta_{(ijkl),(ij)} +\theta_{(ijkm),(ij)} +\theta_{(ijlm),(ij)}
\ea
at an edge $(ij)$, which is shared by the three tetrahedra $(ijkl)$, $(ijkm)$ and $(ijlm)$.

If all the bulk edges $(0i)$ are timelike and large, we know from (\ref{Volhom}) that all the boundary tetrahedra of the initial $5-1$ Pachner move configuration have to be spacelike. The 3D deficit angles $\epsilon_{(ij)}$ are therefore real, and we obtain a real leading-order term for the 3D Regge action, which agrees with $1/2$ of the Regge action for the spacelike boundary triangulation 
\ba
S^{5-1} =-\frac{\sqrt{\lambda}}{2} S^{E,3D} + \mathcal{O}(\lambda^0)\,,
\ea 
where we define the Euclidean Regge action by
\be 
S^{E,3D} = -\sum_{1\leq i<j\leq 5}\sqrt{s_{ij}} \epsilon_{(ij)}\,. \label{ReggeEuclidean3D51}
\ee

\begin{figure}[t]
	\centering
	\includegraphics[width=.48\textwidth]{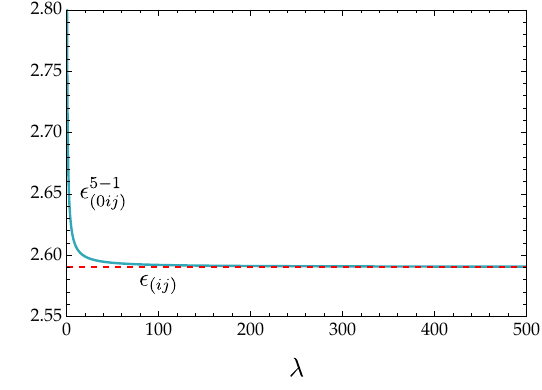}
	\hfill
	\includegraphics[width=.49\textwidth]{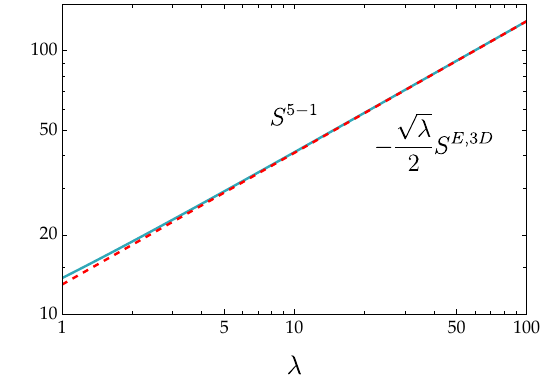}
	\caption{\label{Fig:51BdrySpacelike} Here we consider a $5-1$ Pachner move configuration with spacelike equilateral boundary edges  ($s_{ij}=1$) and timelike bulk edges $s_{0i}=-\lambda$. Left: The bulk deficit angle $\epsilon^{5-1}_{(0ij)}$ (teal curve), and 3D deficit angles $\epsilon_{(ij)}$ in (\ref{eq:3Ddeficit}) (red dashed line) as functions of $\lambda$. Right: The log-log plot of the exact Regge action $S^{5-1}$ (teal curve), and the asymptotic approximation $-\frac{\sqrt{\lambda}}{2}S^{E,3D}$ defined in (\ref{ReggeEuclidean3D51}) (red dashed curve) as functions of $\lambda$.}
\end{figure}

In the case that all the bulk edges $(0i)$ are spacelike and large, the boundary tetrahedra need to be timelike, in order for the triangle inequalities to be satisfied.  Therefore, we have a three-dimensional boundary triangulation with Lorentzian spacetime signature, and can write for the 4D action
\ba
S^{5-1} =\frac{\sqrt{\lambda}}{2} S^{L,3D} + \mathcal{O}(\lambda^0)\,,
\ea
where the Lorentzian 3D action for the boundary of the 4-simplex $(12345)$ is given by
\be 
S^{L,3D} = -\imath \sum_{1\leq i<j\leq 5}\sqrt{s_{ij}} \epsilon_{(ij)} \, . \label{ReggeLorentzian3D51}
\ee 
The bulk triangle $(0ij)$ is light-cone regular with respect to the 4D triangulation, if and only if the edge $(ij)$ is light-cone regular with respect to the three-dimensional triangulation of the boundary. Similarly, we find a real leading-order term for the 4D Regge action, if the Regge action for the three-dimensional boundary is real. 


~\\
{\it Examples:} 
\begin{itemize}
\item Consider an equilateral 4-simplex $(12345)$, with all squared edge lengths equal to $s_{ij}=1$. We subdivide this 4-simplex into five 4-simplices by inserting a bulk vertex $(0)$ and setting the length square of these bulk edges to $s_{0i}=-\lambda$. For large $\lambda$, the quotient $S^{5-1}/\sqrt{\lambda}$ reproduces $10(2\pi-\arccos(1/3))/2=12.9515$, which is $-S^{E,3D}/2$ for the boundary triangulation. All bulk triangles are timelike and therefore light-cone regular. Figure~\ref{Fig:51BdrySpacelike} compares the exact 4D Regge action to its leading-order asymptotics.
\item Consider a $5-1$ Pachner move configuration with boundary squared edge lengths given by $s_{12}=-3$ and $s_{34}=s_{35}=s_{45}=1$, as well as $s_{kl}=-1/5$ for $k=1,2$ and $l=3,4,5$. We set the bulk squared edge lengths to $s_{0i}=+\lambda$ with $i=1,\ldots,5$. The quotient $S^{5-1}/\sqrt{\lambda}$ asymptotes to $4.24276$, which is indeed 1/2 of the Lorentzian 3D Regge action for the boundary triangulation, as shown in Figure~\ref{Fig:51+++++}. Here, we have spacelike bulk triangles $(034)$, $(035)$, and $(045)$, which are light-cone regular.

\begin{figure}[t]
	\centering
	\includegraphics[width=.48\textwidth]{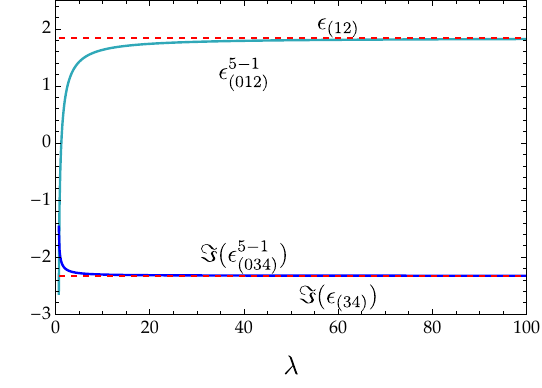}
	\hfill
	\includegraphics[width=.49\textwidth]{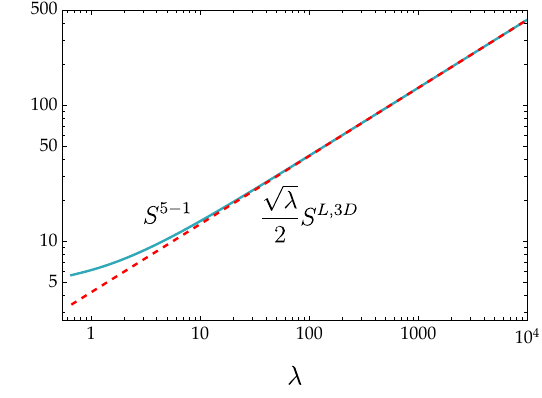}
	\caption{\label{Fig:51+++++} Here we consider a $5-1$ Pachner move configuration with $s_{12}=-3$, $s_{34}=s_{35}=s_{45}=1$, $s_{kl}=-1/5$ for $k=1,2$ and $l=3,4,5$, and spacelike bulk edges $s_{0i}=\lambda$, which are sent to infinity. Left panel: The bulk deficit angle $\epsilon^{5-1}_{(012)}$ (teal curve), the imaginary part of $\epsilon^{5-1}_{(034)}$ (blue curve), 3D deficit angles $\epsilon_{(12)}$ (red dashed line), and the imaginary part of $\epsilon_{(34)}$ (red dashed line) in (\ref{eq:3Ddeficit}) as functions of $\lambda$. Right panel: The exact Regge action $S^{5-1}$ (teal curve), and the asymptotic approximation $\frac{\sqrt{\lambda}}{2}S^{L,3D}$ defined in (\ref{ReggeLorentzian3D51}) (red dashed curve) as functions of $\lambda\in(\frac{79}{120},10^4]$. }
\end{figure}

\item We change the length assignments above by only switching the sign for the $s_{kl}$, which are now $s_{kl}=+1/5$ for $k=1,2$ and $l=3,4,5$. The quotient $S^{5-1}/\sqrt{\lambda}$ now asymptotes to $-0.788443-8.42978\imath$, which is indeed 1/2 of the Lorentzian 3D Regge action for the boundary triangulation. Here, in the 4D triangulation, we have light-cone irregular bulk triangles $(0ik)$, for $i=1,2$ and $k=3,4,5$. The result is shown in Figure~\ref{Fig:51+++++2}.

\begin{figure}[t]
	\centering
	\includegraphics[width=.48\textwidth]{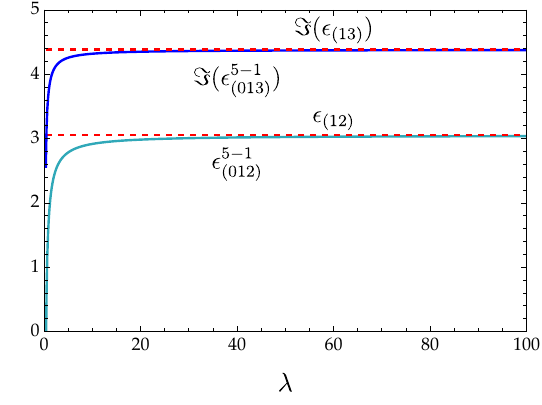}
	\hfill
	\includegraphics[width=.49\textwidth]{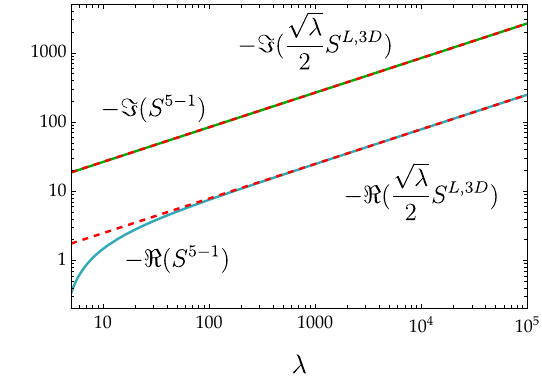}
	\caption{\label{Fig:51+++++2} Here we consider a $5-1$ Pachner move configuration with $s_{12}=-3$, $s_{34}=s_{35}=s_{45}=1$, $s_{kl}=1/5$ for $k=1,2$ and $l=3,4,5$, and spacelike bulk edges $s_{0i}=\lambda$, which are sent to infinity. Left: The bulk deficit angle $\epsilon^{5-1}_{(012)}$ (teal curve), the imaginary part of $\epsilon^{5-1}_{(034)}$ (blue curve), 3D deficit angles $\epsilon_{(12)}$ (red dashed line), and the imaginary part of $\epsilon_{(34)}$ (red dashed line) in (\ref{eq:3Ddeficit}) as functions of $\lambda$. Right: The exact real and imaginary part of Regge action $S^{5-1}$ (teal and green curve), and the asymptotic approximation $\frac{\sqrt{\lambda}}{2}S^{L,3D}$ defined in (\ref{ReggeLorentzian3D51}) (red dashed curve) as functions of $\lambda$. }
\end{figure}

\end{itemize}

~\\	\noindent
{\bf Case $s_{01}=\pm \lambda$  and $s_{0j}=\mp \lambda$ for $j=2,\ldots,5$:}\\\noindent
We have two types of 4-simplices. Firstly, there is the 4-simplex $(02345)$, for which all edges of type $(0i)$ agree in their signature, and for which we have already discussed the behavior of the dihedral angles. Secondly, there are the four 4-simplices $(01ijk)$ with $2\leq i<j<k\leq 5$. For the latter, we have two types of bulk triangles: bulk triangles $(01i)$ and bulk triangles $(0ij)$ with $2\leq i<j\leq 5$.

Consider, for example, the dihedral angle at the triangle $(012)$ in the 4-simplex $(01234)$,
\ba
	\sin(\theta_{(01234),(012)}) \,=\, -\frac{4}{3} \frac{\sqrt{\mathbb{V}_{012}}\sqrt{\mathbb{V}_{01234}}}{\sqrt{\mathbb{V}_{0123}}\sqrt{\mathbb{V}_{0124}}}
 &=& -2 \frac{\sqrt{-\lambda^2}\sqrt{ -\mathbb{V}_{234} \lambda^2 }}  {\sqrt{-s_{23}\lambda^2}\sqrt{-s_{23}\lambda^2}}   \,+\, \mathcal{O}(\lambda^{-1})\nn\\
 &=&\sin(\theta_{(234),(2)}) \,+\, \mathcal{O}(\lambda^{-1}) \, .
 \ea
Similarly, we find $\cos(\theta_{(01234),(012)})=\cos(\theta_{(234),(2)})+\mathcal{O}(\lambda^{-1})$.

For the dihedral angle at the triangle $(023)$ in the 4-simplex $(01234)$, we find
\ba
\sin(\theta_{(01234),(023)}) \,=\, -\frac{4}{3} \frac{\sqrt{\mathbb{V}_{023}}\sqrt{\mathbb{V}_{01234}}}{\sqrt{\mathbb{V}_{0123}}\sqrt{\mathbb{V}_{0234}}}
&=&-\frac{\sqrt{\mp s_{23}\lambda} \sqrt{-\mathbb{V}_{234}\lambda^2  }}{\sqrt{-s_{23}\lambda^2}\sqrt{\mp\mathbb{V}_{234}\lambda }}\,+\, \mathcal{O}(\lambda^{-1}) \nn\\
&=&-1 \,+\, \mathcal{O}(\lambda^{-1}) \, .
\ea
Similarly, we find that $\cos(\theta_{(01234),(023)}) = \mathcal{O}(\lambda^{-1/2})$, such that
\ba\label{77b}\label{eq:Theta01234T023}
\theta_{(01234),(023)}&=&-\pi/2+\mathcal{O}(\lambda^{-1/2})  \, .
\ea

We have already determined the dihedral angles at the triangles $(0ij)$ in the 4-simplex $(02345)$ as
\ba\label{eq:Theta02345T023}
\theta_{(02345),(0ij)}&=&\theta_{(2345),(ij)}\,+\, \mathcal{O}(\lambda^{-1})  \, .
\ea

The dihedral angles at the boundary triangles are at most $\mathcal{O}(\log\lambda)$. Therefore, the boundary deficit angles do not contribute to the leading-order terms in the Regge action.

Next, let us consider the bulk deficit angles. For the deficit angle at $(012)$, we find
\ba
\epsilon_{(012)}^{5-1} &  =&
 2\pi + \theta_{(01234),(012)} +   \theta_{(01235),(012)}+ \theta_{(01245),(012)}\nn \\
 &=&  2\pi + \theta_{(234),(2)} +   \theta_{(235),(2)}+ \theta_{(245),(2)} \,+\, \mathcal{O}(\lambda^{-1})  \, ,
\ea
whereas for the deficit angle at the triangle $(023)$, we find
\ba\label{eq45}
\epsilon_{(023)}^{5-1} &  =&
 2\pi + \theta_{(01234),(023)} +   \theta_{(01235),(023)}+ \theta_{(02345),(023)}\nn \\
&=&\pi + \theta_{(2345),(23)} \,+\, \mathcal{O}(\lambda^{-1/2})  \, .
 \ea

The areas for triangles $(01i)$ with $i\geq 2$ are given by $\mathbb{V}_{01i}=-\lambda^2/4+\mathcal{O}(\lambda^0)$, whereas for the triangles $(0ij)$ with $j>i\geq 2$, the areas are given by $\mathbb{V}_{0ij}=\mp\lambda s_{ij}/4+\mathcal{O}(\lambda^0)$. 

To leading order, we therefore need to consider only the deficit angles at the triangles $(01i)$. The Regge action is then given by
\ba\label{eq:51moveasymptotic2}
S^{5-1} &=& \frac{1}{2} \lambda \sum_{i=2,\ldots,5} \epsilon_{01i} 
\,\,+\,\,\mathcal{O}(\lambda^{1/2}) \nn\\
&=& 2\pi\lambda \,\,+\,\,\mathcal{O}(\lambda^{1/2})\, ,
\ea
where we have again used the fact that the angles in a triangle sum up to $-\pi$. Therefore, we find that the leading-order term in the action is real. This result can be explained by the triangles $(01i)$, which are timelike, and can therefore not be light-cone irregular.  The next-to-leading term scales with $\lambda^{1/2}$.

Next, we determine these subleading $\lambda^{1/2}$ terms. To this end, we note that the next-to-leading-order correction coming from the $(01i)$ triangles is of order ${\cal{O}}(\lambda^0)$, so the $\lambda^{1/2}$ correction comes from the six triangles $(0ij)$ with $5\geq j>i\geq 2$. We determined the associated deficit angles in (\ref{eq45})
\ba\label{eq45b}
\epsilon_{(0ij)}^{5-1} 
&=&\pi + \theta_{(ijkl),(ij)} \,+\, \mathcal{O}(\lambda^{-1/2})  \, ,
 \ea
where $1< i,j,k,l\leq 5$ are pairwise distinct. With $\mathbb{V}_{0ij} = \mp \lambda s_{ij}/4 + \mathcal{O}(\lambda^0)$, the action evaluates to
\ba\label{eq:51moveasymptotic2b}
S^{5-1} 
&=& 2\pi\lambda \,
-\, \frac{\imath}{2}\sum_{2\leq i<j \leq 5}\sqrt{\mp \lambda s_{ij}}\qty(\pi +\theta_{(2345),(ij)})
\,\,+\,\,
\mathcal{O}(\log(\lambda)) \, .
\ea
In the case where $s_{0i}=-\lambda$ for $2\leq i\leq 5$, and $\lambda$ is large, we know from equation (\ref{Volhom}) that the Lorentzian triangle inequalities imply that the tetrahedron $(2345)$ is spacelike. Thus, $s_{ij}\geq 0$, and $\sum_{2\leq i<j \leq 5} \sqrt{s_{ij}} \qty(\pi + \theta_{(2345),(ij)}) = -S^{E,3D}_{(2345)}$, which is minus the Euclidean Regge action for the tetrahedron $(2345)$. Therefore, in the case where four of the bulk edges are timelike, the action is given by
\ba\label{eq:51moveasymptotic2c}
S^{5-1} 
&=& 2\pi\lambda \,
-\, \frac{1}{2} \sqrt{\lambda}\, S^{E,3D}_{(2345)}
\,\,+\,\,
\mathcal{O}(\log(\lambda)) \, .
\ea
Here, there are no light-cone irregular bulk triangles, as all these triangles are timelike. Thus, the leading and next-to-leading order terms in the action, both determined by the deficit angles associated with the bulk triangles, are real.

In the case that $s_{0i}=+\lambda$ for $2\leq i\leq 5$, and $\lambda$ is large, the tetrahedron $(2345)$ has to be timelike. Here, $-\imath\sum_{2\leq i<j \leq 5}\sqrt{ s_{ij}}\qty(\pi +\theta_{(2345),(ij)})=S^{L,3D}_{(2345)}$ evaluates to the Lorentzian action for the tetrahedron $(2345)$, and in this case
\ba\label{eq:51moveasymptotic2d}
S^{5-1} 
&=& 2\pi\lambda \,
+\, \frac{1}{2} \sqrt{\lambda}\, S^{L,3D}_{(2345)}
\,\,+\,\,
\mathcal{O}(\log(\lambda)) \, .
\ea
According to (\ref{Volnonhom1}) applied to the volume of a 4-simplex with one large timelike edge and three large spacelike  edges, the triangles $(ijk)$ with $2\leq i<j<k\leq 5$ have to be spacelike. Thus, the tetrahedron $(2345)$ is timelike, but all its triangles (and therefore edges) are spacelike. According to (\ref{Volhom}), this implies that the bulk triangles $(0ij)$ are spacelike and might be light-cone irregular. Whether this is the case is determined by the 3D dihedral angles at $(ij)$ in the tetrahedron $(2345)$: the triangle $(0ij)$ is irregular if the 3D dihedral angle at $(ij)$ contains more or less than one light cone.
Note that in a timelike tetrahedron with only spacelike triangles, the 3D dihedral angles contain either zero or one light cone. Such tetrahedra also contain at least one edge with a dihedral angle containing zero light cones.~\footnote{
Consider an embedding of the tetrahedron into Minkowskian space. By replacing the Minkowski metric with a Euclidean metric, we obtain a Euclidean tetrahedron. The sum of the dihedral angles in a Euclidean tetrahedron is bounded by $2\pi$ from below and $3\pi$ from above~\cite{Gaddum}. We note that the dihedral angles of a Lorentzian tetrahedron, which has only spacelike triangles, contain either zero or one light cone. Accordingly, the dihedral angles of the associated Euclidean tetrahedron are either smaller or larger than $\pi/2$, and they are always positive. Thus, some of these angles are smaller than $\pi/2$, meaning that some of the Lorentzian angles do not contain a light cone.
} Therefore, there are imaginary contributions to the action $S^{L,3D}_{(2345)}$, and these all have the same negative sign and cannot cancel each other out.

\begin{figure}[t]
	\centering
	\includegraphics[width=.48\textwidth]{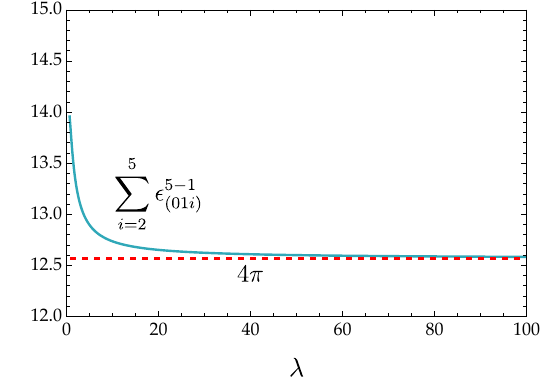}
	\hfill
	\includegraphics[width=.49\textwidth]{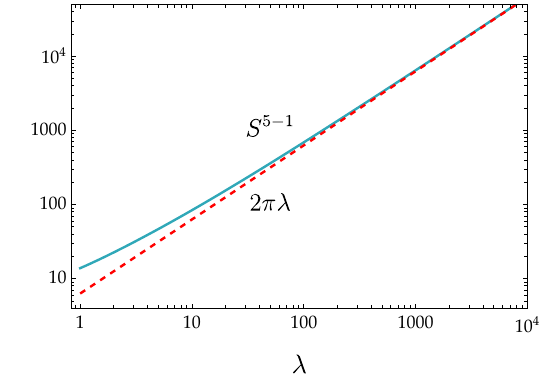}
	\caption{\label{Fig:51moveexple2}  Here, we consider a $5-1$ Pachner move configuration with equilateral boundary edges ($s_{ij}=1$), one spacelike bulk edge, and four timelike bulk edges ($s_{01}=\lambda$,  $s_{02}=s_{03}=s_{04}=s_{05}=-\lambda$). Left: The sum of the bulk deficit angles $\epsilon^{5-1}_{(01i)}, i=2,3,4,5$ (teal curve) and the asymptotic approximation of $4\pi$ (red dashed line) as functions of $\lambda$. Right: The the real part of the exact Regge action $S^{5-1}$ (teal curve) and the leading-order of the asymptotic approximation~\eqref{eq:51moveasymptotic2c} (red dashed curve) as functions of $\lambda$.}
\end{figure} 

\begin{figure}[t]
	\centering
	\includegraphics[width=.48\textwidth]{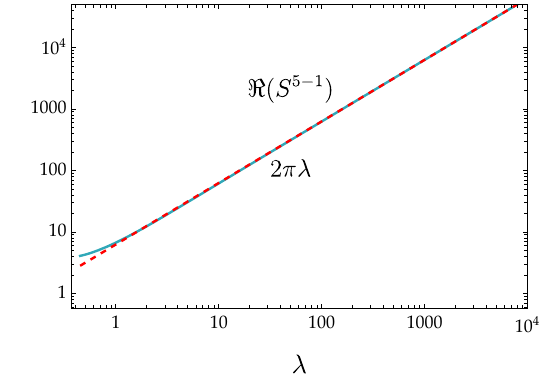}
	\hfill
	\includegraphics[width=.48\textwidth]{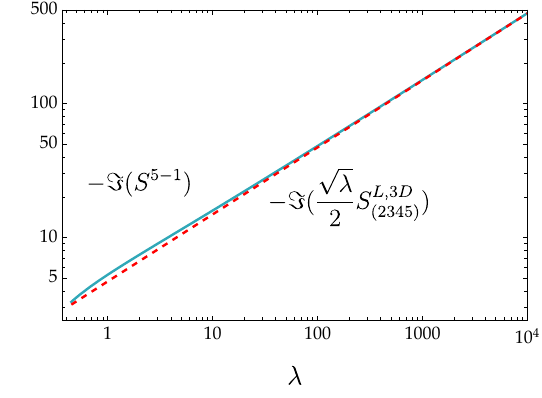}
	\caption{\label{Fig:51move-++++} Here, we consider a $5-1$ Pachner move configuration with boundary squared edge lengths $s_{1i}=1$ for $i=2,\ldots,5$, and $s_{23}=s_{24}=s_{34}=1$ and $s_{25}=s_{35}=s_{45}=3/10$. There is one spacelike bulk edge and four timelike bulk edges $s_{01}=-\lambda$,  $s_{02}=s_{03}=s_{04}=s_{05}=\lambda$. Left: The real part of the exact Regge action $S^{5-1}$ (teal curve) and the real part of the leading-order asymptotic approximation~\eqref{eq:51moveasymptotic2d} (red dashed curve) as functions of $\lambda$. Right: The imaginary part of the exact Regge action $S^{5-1}$ (teal curve) and the imaginary part of the next-to-leading-order asymptotic approximation~\eqref{eq:51moveasymptotic2d} (red dashed curve) as functions of $\lambda$.}
\end{figure}

~\\
{\it Examples:}
\begin{itemize}
\item
We choose all boundary edge lengths squared to be $s_{ij}=1$ and the bulk squared edge lengths to be $s_{01}=\lambda$ and $s_{02}=s_{03}=s_{04}=s_{05}=-\lambda$. An explicit calculation of the Regge action yields a leading-order term of $2\pi\lambda$ and a next-to-leading term of $6(\pi-\arccos{1/3})/2\sqrt{\lambda}=5.7319\sqrt{\lambda}$, which is indeed equal to $-S^{E,3D}_{(2345)}\sqrt{\lambda}/2$. Figure~\ref{Fig:51moveexple2} compares the exact Regge action to its leading asymptotic order.
\item
We choose the following boundary squared edge lengths: $s_{1i}=1$ for $i=2,\ldots,5$ and $s_{23}=s_{24}=s_{34}=1$ and $s_{25}=s_{35}=s_{45}=3/10$. This allows $s_{01}=-\lambda$ and $s_{0i}=+\lambda$ ($i=2,\ldots,5$) with $\lambda$ large.
Computing the exact Regge action for this case and extracting the asymptotics gives $2\pi\lambda -(0.0725118 + 9.42478\imath)\sqrt{\lambda}/2+ {\cal O}(\log\lambda)$. The 3D Lorentzian action for the tetrahedron $(2345)$ evaluates indeed to $-(0.0725118 + 9.42478\imath)$. The action includes an imaginary part of order $\sqrt{\lambda}$, showing that there are light-cone irregular bulk triangles. The corresponding comparison is shown in Figure~\ref{Fig:51move-++++}.
\end{itemize} 

~\\	\noindent
{\bf Case $s_{01}=s_{02}=\pm \lambda$  and $s_{0j}=\mp \lambda$ for $j=3,4,5$:}\\\noindent
Here, there are two types of 4-simplices, firstly the set $\{(01345),(02345)\}$ and secondly the set $\{(012ij),3\leq i<j\leq 5\}$.  For the first set we have one large edge with opposite signature to the other three large edges, and we have already determined the dihedral angles at the bulk triangles. For the second set we have two large edges with opposite signature to the other two large edges.

We know that the area for the bulk triangles with two large edges of opposite signature will scale as $\lambda$, whereas for the bulk triangles with two large edges of the same signature we have a scaling with $\lambda^{1/2}$. The set of triangles whose areas scale with $\lambda$, is given by the bulk triangles $(0ij)$ where $i=1,2$ and $j = 3,4,5$. The second set is given by the bulk triangles $(012)$ and $ (0kl)$ where $k,l = 3,4,5$ with $ k<l$. 

Let us first determine the dihedral angles for the triangles with dominant area scaling. 

We already determined, that, e.g., 
\ba
\theta_{(01345),(013)}&=&\theta_{(345),(3)} \,+\, \mathcal{O}(\lambda^{-1})  \, .
\ea

But for the dihedral angles in e.g. $(01234)$ we find  more complicated expressions
\ba
\sin \theta_{(01234),(013)}&=&
-\frac{4}{3} \frac{\sqrt{\mathbb{V}_{013}}\sqrt{\mathbb{V}_{01234}}}{\sqrt{\mathbb{V}_{0123}}\sqrt{\mathbb{V}_{0134}}}
\nn\\
&=& -6 \frac{ \sqrt{ \left( \frac{\partial}{\partial s_{13}}  +   \frac{\partial}{\partial s_{14}}  +  \frac{\partial}{\partial s_{23}}+\frac{\partial}{\partial s_{24}}     \right)\mathbb{V}_{1234}}}{\sqrt{s_{12}}\sqrt{s_{34}}} \,+\, \mathcal{O}(\lambda^{-1})
 \nn\\
 &=&- \frac{1}{2} \frac{\sqrt{4 s_{12}s_{34}-(s_{13}-s_{14}-s_{23}+s_{24})^2}}{\sqrt{s_{12}}\sqrt{s_{34}}} \,+\, \mathcal{O}(\lambda^{-1})\,,
\ea
and
\ba
\cos \theta_{(01234),(013)}&=&
 \frac{4^2}{\sqrt{\mathbb{V}_{0123}}\sqrt{\mathbb{V}_{0134}}} \frac{\partial}{\partial s_{24}}\mathbb{V}_{01234}\nn\\
 &=&\frac{1}{2}\frac{-s_{13}+s_{14}+s_{23}-s_{24}}{\sqrt{s_{12}}\sqrt{s_{34}}} \,+\, \mathcal{O}(\lambda^{-1}) \, .
\ea
Below we need to know whether $\sin \theta_{(01234),(013)}$ is positive or negative. From Equation (\ref{Vol4dC}) we see that $4 s_{12}s_{34}-(s_{13}-s_{14}-s_{23}+s_{24})^2$ has to be positive, and thus $s_{12}s_{34}$ has to be positive. From  (\ref{Volnonhom1}) we can conclude that the triangle $(345)$ has to be spacelike, and therefore $s_{34}$ positive. Thus, $s_{12}$ has also to be positive, and we see that $\sin \theta_{(01234),(013)}$ is negative.

Now deriving the same expressions for $\theta_{(01234),(014)}$ as for $\theta_{(01234),(013)}$, we notice that 
\ba
\sin \theta_{(01234),(014)}&=&+\sin \theta_{(01234),(013)} \,+\, \mathcal{O}(\lambda^{-1})\,, \q \nn\\
\cos \theta_{(01234),(014)}&=&-\cos \theta_{(01234),(013)}\,+\, \mathcal{O}(\lambda^{-1}) \, .
\ea
Together with the fact that $\sin \theta_{(01234),(013)}<0$, we conclude $\theta_{(01234)(013)}+\theta_{(01234)(014)}=-\pi +\mathcal{O}(\lambda^{-1}) $. 

In general we have 
\ba
\theta_{(012kl)(01k)}+\theta_{(012kl)(01l)}&=&-\pi\,+\, \mathcal{O}(\lambda^{-1}) \, , \nn\\
\theta_{(012kl)(02k)}+\theta_{(012kl)(02l)}&=&-\pi\,+\, \mathcal{O}(\lambda^{-1})  \, ,
\ea
for $3\leq <k<l\leq 5$.

Let us now come to the deficit angles. 
The deficit angle at the triangle $(013)$ is given by
\ba
\epsilon^{5-1}_{013}&=&2\pi + \theta_{(345),(3)} +\theta_{(01234),(013)} +\theta_{(01235),(013)} \,+\, \mathcal{O}(\lambda^{-1}) \, .
\ea
Similarly, we have to consider deficit angles at five additional triangles, that is consider the set of deficit angles at  $\{(013),(014),(015),(023),(024),(025)\}$

For the sum over these deficit angles we obtain 
\ba
\sum_{\substack{i=1,2 \\ j=3,4,5}} \epsilon^{5-1}_{0ij} &=& 12\pi + 2\sum_{k=3}^5\theta_{(345),(k)}+ \sum_{\substack{i=1,2 \\ 3\leq k<l\leq 5}} \left(\theta_{(012kl),(0ik)}+\theta_{(012kl),(0il)}\right) \,+\, \mathcal{O}(\lambda^{-1}) \nn\\
&=& 12\pi \q\q\q -2\pi\q\q\q\q\q\q\q\q\q-6\pi \q\,+\, \mathcal{O}(\lambda^{-1})\nn\\
&=& 4\pi\,\q +\, \mathcal{O}(\lambda^{-1}) \, .
\ea
The signed squared areas of $\{(013),(014),(015),(023),(024),(025)\}$ have all the same large $\lambda$ limit, namely $-\lambda^2/4$. For the leading term of the Regge action we therefore obtain 
\ba\label{eq:51moveasymptotic3}
S^{5-1} &=& \frac{1}{2} \lambda \!\! \sum_{\substack{i=1,2 \\ j=3,4,5}} \epsilon^{5-1}_{0ij} \,\,\,\,+\,\,\,\,  \mathcal{O}(\lambda^{1/2}) \,\,=\,\,
2\pi \lambda \,\,\,+\,\,\,  \mathcal{O}(\lambda^{1/2}) 
\, .
\ea

\begin{figure}[t]
	\centering
	\includegraphics[width=.48\textwidth]{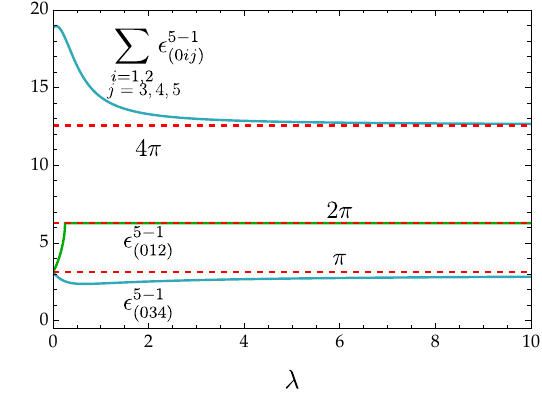}
	\hfill
	\includegraphics[width=.49\textwidth]{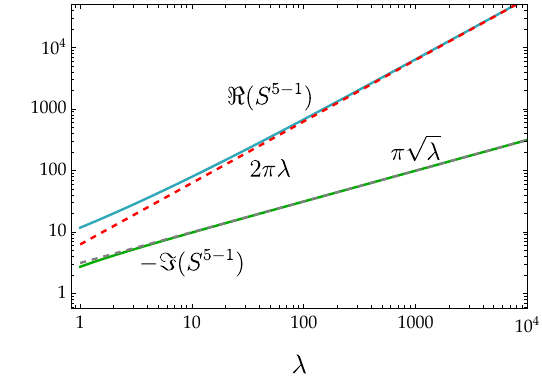}
	\caption{\label{Fig:51moveexple3}  Here we consider a $5-1$ Pachner move configuration with equilateral boundary edges ($s_{ij}=1$), two spacelike bulk edges and three timelike bulk edges ($s_{01}=s_{02}=\lambda$, $s_{03}=s_{04}=s_{05}=-\lambda$). Left: The sum of the bulk deficit angle $\epsilon^{5-1}_{(0ij)},\,i=1,2<j=3,4,5$ (teal curve), the deficit angle $\epsilon^{5-1}_{012}$ at bulk face $(012)$, the deficit angle $\epsilon^{5-1}_{034}$ at bulk face $(034)$ and the the asymptotic approximation $4\pi$, $2\pi$ and $\pi$ (red dashed line) as function of $\lambda$. Right: The real and imaginary part of the exact Regge action $S^{5-1}$ (teal curve and green curve), and the asymptotic approximation~\eqref{eq:51moveasymptotic3} (red and gray dashed curve) as functions of $\lambda$. }
\end{figure}

The next-to-leading-order terms of order $\lambda^{1/2}$ arise solely from the triangles $\{(012),(034),(035),(045)\}$ with two large bulk edges of the same signature. As before, the boundary deficit angles are at most of order ${\cal O}(\log \lambda)$, and the next-to-leading-order coming from the triangles $\{(013),(014),(015),(023),(024),(025)\}$ is of order ${\cal O}(\lambda^{0})$.

The deficit angles associated to the triangles $\{(012),(034),(035),(045)\}$ are given by
\ba
	\epsilon_{(012)}^{5-1} &  =&
	2\pi + \theta_{(01234),(012)} +   \theta_{(01235),(012)}+ \theta_{(01245),(012)}\,,\nn\\
	\epsilon_{(0kl)}^{5-1} &  =&
	2\pi + \theta_{(012kl),(0kl)} +   \theta_{(01klm),(0kl)}+ \theta_{(02klm),(0kl)}\,,
	\ea
with $3\leq k,l,m \leq 5$ and $k,l,m$ pairwise different.
All the dihedral angles in the first line are associated to a bulk triangle with both bulk edges of equal signature in a 4-simplex whose other two bulk edges have opposite signature. We still need to compute these. In the second line, the first dihedral angle is also of this type, whereas for the other two we can use the result~\eqref{eq:Theta01234T023} to conclude that each of these amounts to $-\pi/2 + \mathcal{O}(\lambda^{-1/2})$.
	For the dihedral angles of type $\theta_{(01234),(012)}$ we have
	\ba
	\sin \theta_{(01234),(012)} &=& -\frac{4}{3}\frac{\sqrt{\mathbb{V}_{012}}\sqrt{\mathbb{V}_{01234}}}{\sqrt{\mathbb{V}_{0123}}\sqrt{\mathbb{V}_{0124}}} \nn\\
	&=& -\frac{4}{3} \frac{\sqrt{\pm \frac{1}{4}s_{12}\lambda}\sqrt{-\frac{\lambda^2}{2^4} \qty(\pdv{}{s_{13}} + \pdv{}{s_{14}} + \pdv{}{s_{23}} + \pdv{}{s_{24}})\mathbb{V}_{1234}}}{\sqrt{-\frac{1}{36}s_{12}\lambda^2}\sqrt{-\frac{1}{36}s_{12}\lambda^2}} + \mathcal{O}(\lambda^{-3/2})\,.
	\ea
	Therefore this type of dihedral angle goes to zero in the limit $\lambda\to \infty$. Thus,  the deficit angles above are given by
	\ba
	\epsilon_{(012)}^{5-1} &  =&
	2\pi +\mathcal{O}(\lambda^{-1/2})\,,\nn\\
	\epsilon_{(0kl)}^{5-1} &  =&
	\pi + \mathcal{O}(\lambda^{-1/2})\,.
	\ea
The leading order of the area square for the triangle $(012)$  is given by $\pm\tfrac{1}{4}\lambda s_{12}$ and the leading order for the triangles $(0kl)$ (with $3\leq k<l\leq 5$) is $\mp\tfrac{1}{4}\lambda s_{kl}$. We thus obtain for the action
\ba\label{eq:51moveasymptotic3b}
S^{5-1} &=& 
2\pi \lambda  \,\,
-\imath  \lambda^{1/2}\left( \pi \sqrt{\pm s_{12}} + \frac{\pi}{2} \sum_{3\leq k<l\leq 5} \sqrt{\mp s_{kl}}\right)
\,\,\,+\,\,\,  \mathcal{O}(\log \lambda)
\, .
\ea

Let us first consider the case where $s_{01}=s_{02} =+\lambda$ and $s_{0j}=-\lambda$ for $j=3,4,5$.  The tetrahedron $(0123)$ has to be timelike, and as it has two large spacelike edges $(01)$ and $(02)$ and one large timelike edge $(03)$, $s_{12}$ has to be spacelike, see (\ref{Volnonhom1}). The triangles $(0kl)$ with $3\leq k<l\leq 5$ are timelike, according to Equation~\eqref{Vol2dA}, the edges $(kl)$ have therefore to be spacelike. Thus the first term in the brackets in (\ref{eq:51moveasymptotic3b}) is real, whereas the second one is imaginary. 

In the second case, where $s_{01}=s_{02} =-\lambda$ and $s_{0j}=+\lambda$, one also finds that $(01)$ has to be spacelike (due to the triangle $(012)$ being timelike and (\ref{Vol2dA})) and that the edges $(kl)$ have to be spacelike (due to the tetrahedra $(01kl)$ being timelike and~\eqref{Volnonhom1}). Thus the first term in the brackets in (\ref{eq:51moveasymptotic3b}) is imaginary, whereas the second one is real.  

In summary, the action includes in both cases imaginary terms of order $\lambda^{1/2}$. These arise from the spacelike bulk triangles, which are all light-cone irregular and of yarmulke type.

~\\
{\it Examples:}
\begin{itemize}
\item We choose all boundary squared edge lengths to be $s_{ij}=1$ and the bulk squared edge length to be $s_{01}=s_{02}=\lambda$ and $s_{03}=s_{04}=s_{05}=-\lambda$. An explicit computation of the Regge action leads indeed to the asymptotic expression (\ref{eq:51moveasymptotic3b}) and thus also yields the imaginary part $-\imath \pi\lambda^{1/2}+{\cal O}(\log \lambda)$. See Figure~\ref{Fig:51moveexple3} for a comparison of the real part of the action with the leading-order asymptotics.
\item With the same boundary data we can also choose  $s_{01}=s_{02}=-\lambda$ and $s_{03}=s_{04}=s_{05}=+\lambda$. The explicit computation of the Regge action again conforms with the asymptotic expression (\ref{eq:51moveasymptotic3b}).
\end{itemize}

Allowing for a relabeling of the vertices, we have covered all possible cases with different number of spacelike and timelike bulk edges. A summary of the various cases can be found below.

\subsection{Summary of asymptotic behaviour of the Regge action}

Here we summarize the asymptotic behaviour of the Regge action for the different Pachner moves and the various choices of signature for the bulk edges:
\begin{itemize}
\item $4-2$: Case $s_{01}=-\lambda$:
\ba
 S^{4-2} &=& \,=\,   \pi  \lambda + \mathcal{O}(\log \lambda)
\ea
\item $4-2$: Case $s_{01}=\lambda$:
\ba
 S^{4-2} &=& \,=\,   \pi  \lambda + \mathcal{O}(\log \lambda)
\ea
\item $5-1$: Case $s_{0i}=-\lambda$  all $i=1,\ldots,5$:
\ba
S^{5-1} =-\frac{\sqrt{\lambda}}{2} S^{E,3D} + \mathcal{O}(\lambda^0)
\ea
\item $5-1$: Case $s_{0i}=+\lambda$  all $i=1,\ldots,5$:
\ba
S^{5-1} =\frac{\sqrt{\lambda}}{2} S^{L,3D} + \mathcal{O}(\lambda^0)
\ea
\item $5-1$: Case $s_{01}=+\lambda$  and $s_{0j}=- \lambda$ for $j=2,\ldots,5$:
\ba
S^{5-1} 
&=& 2\pi\lambda \,
-\, \frac{1}{2} \sqrt{\lambda}\, S^{E,3D}_{(2345)}
\,\,+\,\,
\mathcal{O}(\log(\lambda)) 
\ea
\item $5-1$: Case $s_{01}=-\lambda$  and $s_{0j}=+ \lambda$ for $j=2,\ldots,5$:
\ba
S^{5-1} 
&=& 2\pi\lambda \,
+\, \frac{1}{2} \sqrt{\lambda}\, S^{L,3D}_{(2345)}
\,\,+\,\,
\mathcal{O}(\log(\lambda)) 
\ea
\item
$5-1$ Case $s_{01}=s_{02}=+ \lambda$  and $s_{0j}=- \lambda$ for $j=3,4,5$:
\ba
S^{5-1} &=& 
2\pi \lambda  \,\,
-\imath  \lambda^{1/2}\left( \pi \sqrt{ |s_{12}|} + \frac{\imath \pi}{2} \sum_{3\leq k<l\leq 5} \sqrt{ |s_{kl}}|\right)
\,\,\,+\,\,\,  \mathcal{O}(\log \lambda)
\ea
\item
$5-1$: Case $s_{01}=s_{02}=- \lambda$  and $s_{0j}=+\lambda$ for $j=3,4,5$:
\ba
S^{5-1} &=& 
2\pi \lambda  \,\,
-\imath  \lambda^{1/2}\left( \imath\pi \sqrt{ |s_{12}|} + \frac{\pi}{2} \sum_{3\leq k<l\leq 5} \sqrt{| s_{kl}|}\right)
\,\,\,+\,\,\,  \mathcal{O}(\log \lambda)
\ea
\end{itemize}

We see that the terms of order $\lambda$ do not depend on the boundary data, and are associated to timelike bulk triangles. The terms of order $\sqrt{\lambda}$ do depend on the boundary data. In the cases where the bulk edges have not all the same signature, and where there are spacelike bulk triangles, there are also light-cone irregular bulk triangles.

\section{Finite expectation values for spine and spine configurations}\label{Sec:PathIntegralAsymptotics}

Spike and spine configurations involving bulk edges which can have  arbitrarily large lengths pose a challenge to the convergence of the gravitational path integral. 
In Euclidean Regge quantum gravity, spike configurations are highly problematic. It has been shown for two-dimensional Regge calculus, that (for a measure that does not suppress edge lengths exponentially) expectation values for sufficiently high powers of the bulk lengths in fact diverge~\cite{Ambjorn:1997ub}. In higher dimensions, the $5-1$ and $4-1$ Pachner move in four and three spacetime dimensions, respectively, isolate the conformal degree of freedom~\cite{Dittrich:2011vz,Borissova:2023izx}. This mode comes with the ``wrong" sign~\cite{Gibbons:1978ac} and leads to an exponential enhancement of these spike configurations in the Euclidean path integral. Due to an infinite integration range, this leads to divergences. 

In contrast,  the oscillatory nature of the path integral for Lorentzian Regge calculus might avoid the conformal factor problem. This has been shown to be the case for the theory linearized around a flat background in~\cite{Borissova:2023izx}.  Moreover, in three spacetime dimensions, the recent work~\cite{Borissova:2024pfq} showed that also in the full theory, arbitrary expectation values of powers of bulk edge squares in (light-cone regular $4-1$ and $3-2$ Pachner moves) spike and spine configurations remain finite. Here we will consider the four-dimensional Lorentzian Regge path integral and investigate the convergence for spine and spike configurations appearing in $4-2$ and $5-1$ Pachner moves, respectively.

For the Regge path integral we need to specify a measure. Whereas in three-dimensional Regge calculus there exists a unique local measure which renders the  partition function triangulation-invariant to one-loop order~\cite{Dittrich:2011vz,Borissova:2023izx}, there is no such local measure available for the four-dimensional theory~\cite{Dittrich:2014rha}. For the $5-1$ and $4-2$ moves (for which the Regge action is invariant), it is possible to find a measure which leads to invariance modulo a non-local factor, which does not factorize over the simplices. This measure is given by~\cite{Dittrich:2011vz,Dittrich:2014rha,Borissova:2023izx}
\be
\mathcal{D}s_e =  \frac{1}{\prod_{e\subset \text{bdry}}\sqrt{\sqrt{96 \pi }}} \frac{1}{\prod_{e\subset \text{bulk}}\sqrt{\sqrt{96 \pi }}} \frac{1}{\prod_{\sigma}\abs{\mathbb{V}_\sigma}^{\frac{1}{4}}} \prod_{e\subset \text{bulk}} \dd{s}_e\,,
\ee
where $\sigma$ denotes the 4-simplices in the triangulation. 
We note that we can apply the asymptotic expansion formulae (\ref{Volhom}), (\ref{Volnonhom1}) and (\ref{Vol4dC}) to the squared volumes $\mathbb{V}_\sigma$, and obtain a measure that includes inverse powers of the bulk edge lengths.  

In the following we will assume that the dependence of the measure on the bulk length squared variables in the asymptotic regime is given by a fractional positive or negative power. As our primary goal is to argue for the finiteness of expectation values for spine and spike configurations, we only need to consider the asymptotic regime of large bulk edge squared variables.

Here we will only focus on  spine and spike configurations with light-cone regular triangles and therefore real contributions to the Regge action. As discussed in Section~\ref{Sec:CRegge},  light-cone irregular triangles lead to branch cuts for the Regge action and imaginary terms. The sign for these imaginary terms changes if one crosses the branch cuts. Choosing the path integral contour such that it goes along the suppressing side, one can exponentially suppress these configurations.
To study the convergence properties of the path integral, we therefore only need to consider light cone regular configurations.

\subsection{$4-2$ Pachner move}

For the $4-2$ Pachner move,  the path integral amounts to an integral over one bulk squared edge variable, $s_{01}=\pm \lambda$, $\lambda>0$. We found previously that, independent of the timelike or spacelike nature of the bulk edge, the asymptotic expression for the Regge action is real and given by
\be
S^{4-2} = \pi \lambda + \mathcal{O}\qty(\log \lambda)\,.
\ee
To investigate the convergence of expectation values of powers of the squared length  $s_{01}^n=(\pm)^n\lambda^n$, we use the above asymptotic expression for the Regge action and combine it with our assumption for the asymptotics of the measure $\mu \propto \lambda^M$, where $M$ can be any positive or negative fraction. Setting $m=n + M$, and dropping a sign $(\pm)^n$ we thus have to consider integrals of the type
\be
\tilde{\mathcal{I}}_{4-2}(m,c) = \int_c^\infty \dd{\lambda} \lambda^m e^{\imath \pi  \lambda}\,,
\ee
where $c>0$ is a sufficiently large positive constant which ensures that we integrate over configurations where the asymptotic approximation is valid. Introducing a regulator $\epsilon>0$ allows us to write the previous integral as
\be\label{eq:Integral42}
\tilde{\mathcal{I}}_{4-2}(m,c) = \lim_{\epsilon\to 0} \int_c^\infty \dd{\lambda}  \lambda^m e^{\qty(\imath \pi - \epsilon)\lambda} = c^{m+1} E_{-m}\qty(- \imath \pi c)\,,
\ee
where $E_n(z)\equiv \int_1^\infty \dd{t} t^{-n}e^{-zt}$ is the exponential integral function. \eqref{eq:Integral42} is finite for any $m\in \mathbb{R}$.

We will use the full action for a numerical computation of the expectation value, and define
\begin{equation}
\label{eq:exp42full}
    \mathcal{E}_{4-2}(m,c)=\int_{c}^{\infty} \mathrm{d}\lambda\,\lambda^m e^{\imath S^{4-2}(\lambda)}\, .
\end{equation}
The lowest possible value for $c$ is determined by the generalized triangle inequalities, but to show finiteness of the expectation values we can also choose a larger $c$.
This integral cannot be solved analytically, due to its highly oscillatory behaviour and the complicated form of the full Regge action. However, as in~\cite{Dittrich:2023rcr,Borissova:2024pfq}, to deal with the unbounded integral we employ series-acceleration methods, like Wynn's epsilon algorithm or iterated Shank's transforms, see~\cite{Schmidt, Shanks, Wynn} and~\cite{WenigerReview} for a review. These methods have been shown to be efficient tools when evaluating path integrals and expectation values~\cite{Dittrich:2023rcr}. The algorithms are particularly efficient if we choose the lower bound $c$ for the integral such that the asymptotic form of the action holds approximately. 

\begin{figure}[t]
	\centering
	\includegraphics[width=.48\textwidth]{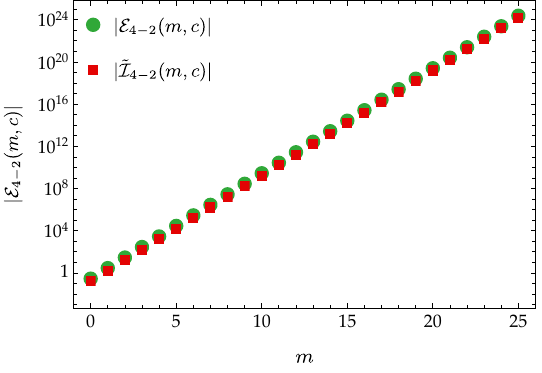}
	\caption{\label{Fig:Wynn42} Comparison of the analytical approximation of the expectation values~\eqref{eq:Integral42} with the full numerical result~\eqref{eq:exp42full} for a configuration with $s_{ij}=2$ (for $2\leq i\leq j\leq 5$) and $s_{0i}=s_{1j}=1/4$ (for $2\leq i\leq 5$) and a spacelike bulk edge, see also Figure~\ref{Fig:42spacelikebulk}. For the numerical evaluation, we employ the Wynn algorithm, and we use $c=10$ for both integrals. The integration over the full action is finite and of the same order of magnitude as the integration over the asymptotic approximation of the action. 
 }
\end{figure}

In Figure~\ref{Fig:Wynn42} we show the comparison of the  expectation value $\mathcal{E}_{4-2}(m,c)$~\eqref{eq:exp42full} defined with the full action with the analytical approximation  $\tilde{\mathcal{I}}_{4-2}(m,c)$~\eqref{eq:Integral42} for the same configuration as displayed in Figure~\ref{Fig:42spacelikebulk}, and for $c=10$. At this lower bound of the integration, the full action $S^{4-2}$ deviates from the asymptotic approximation by almost $40\%$. This also results in a deviation between the full and approximated expectation values of about $30\%$, which however decreases when increasing the power $m$. This is expected, since for larger $m$, the regime of large $\lambda$, where the integrands of both expressions agree better, contributes more. Overall we find stable numerical results for $\mathcal{E}_{4-2}(m,c)$, which remain finite for all tested $m\in[0,25]$.

\subsection{$5-1$ Pachner move}

The $5-1$ Pachner move configuration can be obtained by subdividing a simplex $(12345)$ via a placement of a vertex $(0)$ inside this simplex and by subsequently connecting all boundary vertices with the bulk vertex. In a path integral for such a configuration one would therefore have to integrate over five length squared variables $s_{0i}$, $i=1,\dots, 5$. As discussed previously, we will only consider the path integral for configurations which have only light cone regular bulk triangles in the asymptotic regime.  We thus only need to consider the cases where $(a)$ the signature of all bulk edges is the same and $(b)$ one bulk edge is timelike and the remaining four are spacelike.
The action for these cases is of the form 
\be\label{eq:S51-A}
S^{5-1} = \alpha \lambda + \beta \sqrt{\lambda} + \mathcal{O}\qty(\log \lambda)\,,
\ee
where $\alpha=0$ in the case that the signature for all bulk edges agrees, and $\alpha=2\pi$ in the case that one bulk edge is timelike and the other four are spacelike. $\beta$ is a real constant, which depends on the data of the boundary triangulation. 

As discussed in Section~\ref{Sec:51move}, the $5-1$ configuration features a four-dimensional gauge symmetry, which is a remnant of diffeomorphism symmetry~\cite{Dittrich:2008pw,Bahr:2009ku,Dittrich:2009fb}. 
We therefore must specify how to deal with this gauge symmetry.~\footnote{The squared edge length is not invariant under this gauge symmetry. We will nevertheless consider the corresponding expectation values, in order to be able to compare with statements in previous literature~\cite{Ambjorn:1997ub}. To this end, we can regard the gauge fixing as a form of symmetry reduction. Namely, we will set all bulk edge lengths to be equal. We then compute the expectation value in this symmetry-reduced model. Alternatively, one can construct gauge invariant observables, e.g.~using the relational formalism~\cite{Dittrich:2004cb,Dittrich:2005kc}, and express one edge length in relation to the other four. Assuming that the resulting observable can be approximated by a polynomial function in the edge lengths, the results on the finiteness also apply to this observable.}
Our choice of additive scaling for the bulk edges $s_{0i} = s^0_{0i} \pm \lambda$ already implies a gauge fixing in the asymptotic regime $\abs{s_{0i}} = \lambda$. We thus have to insert a Faddeev-Popov determinant and as a result remain with only a one-dimensional integral over the variable $\lambda$. Along the flat solution, where the gauge orbits can be parametrized explicitly, the Faddeev-Popov determinant can be computed as (see~\cite{Baratin:2006yu,Dittrich:2011vz,Borissova:2023izx} for a similar computation)
\be
\mathcal{F} = 2^4 4! \,V_{(12345)}\,.
\ee
Here $V_{(12345)}$ is the absolute volume of the final simplex $(12345)$, which is independent of the bulk variables and thus independent of $\lambda$. Away from the flat solution, we will assume that the gauge-fixing determinant together with the measure can be approximated asymptotically by some fractional positive or negative power of $\lambda$. We thus have to consider one-dimensional integrals over the variable $\lambda$ with a Regge action of the form~\eqref{eq:S51-A}.

For the case with five bulk edges of equal signature we therefore consider
\be
\tilde{\mathcal{I}}_{5-1}(m,\beta,c) = \int_c^\infty \dd{\lambda} \lambda^m e^{\imath \beta \sqrt{\lambda}} = 2 \int_{\sqrt{c}}^{\infty} \dd{\tilde{\lambda}} \tilde{\lambda}^{2m + 1} e^{\imath \beta \tilde{\lambda}}\,,
\ee
where $c>0$ is again some sufficiently large positive constant such that the asymptotic approximation leads only to a very small correction. The power $m\in \mathbb{R}$ summarizes the contribution from the measure, the Faddeev-Popov determinant, and the insertion of the bulk lengths observables into the path integral.

Similarly as for the $4-2$ Pachner move, these integrals result in
\be
\tilde{\mathcal{I}}_{5-1}(m,\beta,c) = 2\, c^{m+1} E_{-2m -1}\qty(-\imath \beta \sqrt{c})
\ee
and are finite for any value of $m\in \mathbb{R}$.

For the second case with one timelike and four spacelike bulk edges we do have a leading order term of order $\lambda$ and a subleading order term of order $\lambda^{1/2}$ in the action. If we only consider the leading order term the calculation proceeds in the same way as for the $4-2$ move; the only difference is that the leading order term is now $2\pi \lambda$ and not $\pi \lambda$. If we also consider the subleading order term in the action we obtain the integrals
\ba\label{eq:Integral51-2a}
\tilde{\mathcal{I}}_{5-1}(m,\beta,c) &=& \int_c^\infty \dd{\lambda} \lambda^m e^{2 \pi \imath \lambda + \imath \beta \sqrt{\lambda}} \nn\\
&=& \frac{1}{2^m}\int_{\tilde{c}}^\infty \dd{\tilde \lambda}
\left(1-\text{sgn}(b) \frac{b^2}{\sqrt{b^4 + 4 b^2 \tilde \lambda}}\right)\left( b^2 + 2 \tilde \lambda - \text{sgn}(b) \sqrt{b^4+4 b^2 \tilde \lambda}  \right)^m e^{2\pi \imath \tilde \lambda}\, ,\q\q
\ea
where we applied a coordinate transformation $\tilde \lambda = \lambda + b \sqrt{\lambda}$ with $b=\beta/2\pi$. Here $\tilde c=c+b\sqrt{c}$ and we assume $c>0$ and $\sqrt{c}>|b|$, thus $\tilde c>0$. We expand the non-exponential factors in the integral around $\tilde \lambda =\infty$
\ba
\tilde\lambda^m\left(1-\text{sgn}(b) \frac{b^2}{\sqrt{b^4 + 4 b^2 \tilde \lambda}}\right)\left(2+ \frac{b^2}{\tilde\lambda}    - \text{sgn}(b) \sqrt{\frac{b^4}{\tilde\lambda^2}+4 \frac{b^2 }{\tilde \lambda}}  \right)^m = \tilde \lambda^m \sum_{k=0}^\infty a_k(m) \tilde \lambda^{-k/2} \, .
\ea
Given the asymptotic approximation discussed here, one can truncate this series at a sufficiently large but finite $k=K$, so that absolute convergence of the remaining integrals is guaranteed and the errors are sufficiently small:~\footnote{The errors can be estimated using $|\int^\infty_{\tilde{c}} \dd{\tilde \lambda} \tilde \lambda^{m-k/2}  \exp(2\pi \imath \tilde \lambda)|\leq \frac{1}{-m+k/2-1}\tilde c^{m-k/2+1} $ for $m-k/2<-1$, and are thus small if $\tilde c\gg 1$ and $k/2\gg m$. }
\ba
\label{eq:51analyt}
\tilde{\mathcal{I}}_{5-1}(m,\beta,c) &\approx& 
 \frac{1}{2^m} \sum_{k=0}^K a_k(m) \int_{\tilde{c}}^\infty \dd{\tilde \lambda} \tilde \lambda^{m-k/2}  e^{2\pi \imath \tilde \lambda}\nn\\
 &=& \frac{1}{2^m} \sum_{k=0}^K a_k(m)\, \tilde c^{m-k/2+1} E_{-m+k/2}(-\imath 2\pi \tilde c) \, .
\ea
The $E_{-m+k/2}(-\imath 2\pi \tilde c)$ are finite for real $m$ and $k$. We therefore conclude that the expectation values computed with the asymptotic form of the action are finite.

\begin{figure}[t]
	\centering
	\includegraphics[width=.48\textwidth]{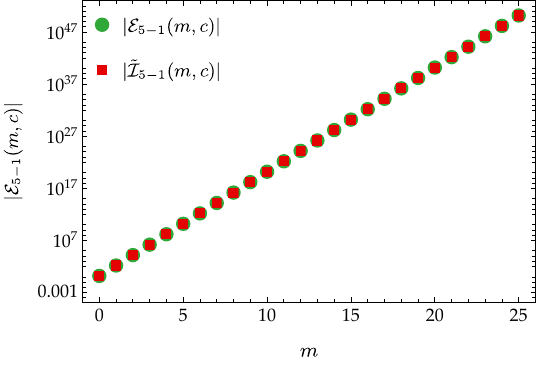}
	\hfill
	\includegraphics[width=.48\textwidth]{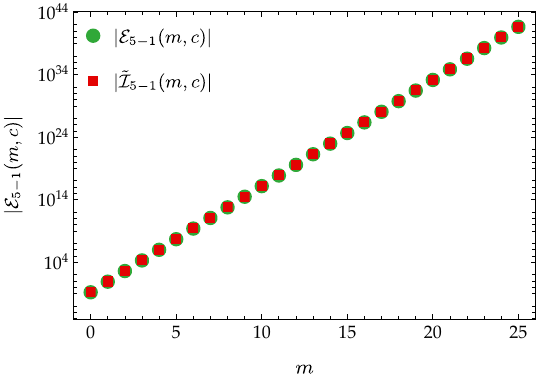}
	\caption{\label{Fig:Wynn51} Comparison of the analytical approximation of the expectation values~\eqref{eq:51analyt} with the full numerical result~\eqref{eq:51movefull} for the cases that $(a)$ (left panel) all bulk edges are timelike and $(b)$ (right panel) one bulk edge is timelike and the remaining four are spacelike.  The integration over the full action is finite, and agrees quantitatively with the integration over the asymptotic limit of the action, for the chosen value of the lower integration bound $c$.}
\end{figure}

Similar to the $4-2$ move, we will now go one step further and compute the expectation values based on the full action. Again, we will employ series-acceleration methods to numerically evaluate the following integrals:
\begin{equation}
\label{eq:51movefull}
    \mathcal{E}^{(j)}_{5-1}(m,c)=\int_{c}^{\infty} \mathrm{d}\lambda\,\lambda^m e^{\imath S^{5-1}_{(j)}(\lambda)}\,,
\end{equation}
where $j={a,b}$ distinguishes the two cases $(a)$ the signature of all bulk edges is the same and $(b)$ one bulk edge is timelike and the remaining four are spacelike. As for the $4-2$ move, the lower integration limit $c$ is bounded from below by the generalized triangle inequalities. But to show finiteness of the expectation values we can choose a larger $c$, and will do so in order to obtain faster convergence for the series acceleration algorithm.  In Figure~\ref{Fig:Wynn51} we compare the full result to the analytic approximation for both cases $(a)$ and $(b)$. For the homogeneous case $(a)$ (left panel), we choose $c=100$, at which the asymptotic approximation to the action deviates from the full action by about $5\cdot10^{-4}$, see also Figure~\ref{Fig:51BdrySpacelike}. The resulting deviation between the approximated and full expectation values is on the per-mil level, and decreases for larger values of $m$. For the inhomogeneous case $(b)$ (right panel), we choose $c=50$, at which the asymptotic approximation to the action deviates from the full action by about $15\%$, see also Figure~\ref{Fig:51moveexple2}. This results in a deviation of the full and approximated expectation values of the order of a few percent, which decreases for increasing $m$. For the comparison of approximated and full result, we neglected the sub-leading term $\beta$ in the analytic expression~\eqref{eq:51analyt}. As for the $4-2$ move, we conclude that the expectation values of the bulk edge to some power $m$ remains finite. Furthermore, these expectation values can be computed numerically, using series-acceleration methods, or analytically, by approximating the action with its asymptotic behaviour for large bulk edges.

\section{Discussion}\label{discussion}

In this paper we investigated spike and spine configurations which appear in simplicical approaches to quantum gravity, such as Regge calculus and spin foams. Such configurations allow for infinitely large bulk edges, and also an infinitely large four-volume, despite keeping the boundary geometry fixed. This illustrates one of the difficulties to define a notion of scale in quantum gravity~\cite{Dittrich:2014ala,Asante:2022dnj,Carrozza:2016vsq}. Due to their unbounded support, spike and spine configurations could potentially dominate the path integral or even lead to divergences. It is therefore important to investigate these configurations and to understand their role in the quantum gravitational path integral. The results of this paper include:
\begin{itemize}
\item The Regge action is a complicated non-polynomial function of the edge lengths, making the extraction of physical properties difficult.
Here we determined  much simpler asymptotic forms for the Regge action for spikes and spines appearing in $5-1$ and $4-2$ Pachner moves, respectively.  In some cases, these simplifications involve a type of dimensional reduction. For the $4-2$ as well as the $5-1$ move, we considered all possible choices for the signature of the bulk edges. In all cases, except for two (which are the $5-1$ move configurations with all edges either spacelike or timelike), we found that the leading-order term of the Regge action is of order $\lambda^1$, where $\lambda$ denotes the absolute value of the squared edge length. The  $\lambda^1$ coefficient is independent of the boundary data. In the two remaining cases, the leading order term is of order $\lambda^{1/2}$. Such terms also appear as subleading terms for the $5-1$ configurations with bulk edges of mixed signature. The $\lambda^{1/2}$ coefficients do depend on the boundary data. 

 The techniques employed here can be generalized to other configurations containing large edge lengths.  Section~\ref{GentoN} already considered a generalization of the $4-2$ Pachner move to configurations of $N$ 4-simplices sharing one large edge. We can expect simplifications of the Regge action along the lines observed in this work, as long as all involved 4-simplices contain a mixture of large and small edges. 
\item We found that in some cases the asymptotic regime exhibits light cone-irregular bulk triangles. This includes all $5-1$ move configurations with inhomogeneous signature for the bulk edges and with spacelike bulk triangles. The situation is however different from the three-dimensional case, where large spacelike bulk edges in the $3-2$ and $4-1$ configurations always lead to light-cone irregular hinges. We found an example for a $5-1$ move configuration with only spacelike bulk edges, which has light-cone regular spacelike bulk triangles. Changing the boundary data, we also found a configuration with light-cone irregular triangles. 
\item In all but one of the cases where the asymptotic regime is light-cone irregular, the light-cone irregular bulk deficit angles are of yarmulke type. That is, the angles include less than two light cones. The only case where we might obtain irregular deficit angles of trouser type, is in the case of the $5-1$ move configuration with only spacelike bulk edges. In this case a bulk triangle is light cone irregular if its boundary edge is light-cone irregular with respect to the Lorentzian boundary triangulation. So far we cannot exclude that trouser type light-cone irregularities might appear in such a Lorentzian boundary triangulation. It would be helpful to obtain a deeper geometric understanding of the light-cone irregular configurations, e.g., by considering how null geodesics traverse such configurations~\cite{Dittrich:2005sy}.

As the revealed light-cone irregular regimes appear for arbitrary large edge lengths, these regimes can only be included in a path integral if the path integral contour is chosen to go along the suppressing side of the branch cuts.  

\item  Our methods allow us to decide which types of asymptotic regimes are allowed by the generalized triangle inequalities and which are forbidden. For instance, a $5-1$ move configuration with spacelike boundary tetrahedra and large spacelike bulk edges is not allowed. But this is the one case, on which numerical efforts of the spin foam community  have been concentrated: the latest result indicates that the associated state sum is finite (with a specific choice of measure)~\cite{Dona:2023myv}.  Given that for this case there are no underlying Regge geometries with arbitrarily large edges, this result may not be surprising. We point out, that there are however many other types of $5-1$ Pachner move configurations, involving timelike tetrahedra, timelike triangles, or timelike edges where the triangle inequalities do allow for large bulk edges. These configurations also need to be investigated, and we cannot draw a conclusion regarding finiteness from the one case which has been investigated so far.

\item The found asymptotic expressions allow us to obtain estimations for the path integral involving the corresponding spike and spine configurations. One can interpret $5-1$ and $4-2$ moves as coarse graining moves~\cite{Dittrich:2013xwa}, and  estimations for the corresponding path integrals can help in understanding renormalization properties of Lorentzian Regge calulus or of spin foams~\cite{Dittrich:2014ala,Asante:2022dnj}.

\item We showed not only that the Lorentzian Regge path integral associated to $5-1$ and $4-2$ configurations is finite, but also that this holds for arbitrary powers of length expectation values and for a large class of measures. This is in sharp contrast to the Euclidean quantum gravity version of Regge calculus. One can show that the $5-1$ spike configurations isolate the conformal factor in the gravitational action~\cite{Dittrich:2011vz}, leading to an exponential enhancement of these configurations.  Thus, without rotating the sign of the conformal factor, such configurations lead to divergences. 

We established finiteness using the asymptotic expressions for the action and an analytical calculation. We also performed the calculation using the exact action and the Wynn algorithm~\cite{Wynn} for numerically dealing with oscillating integrals. The integrals are also finite for cases where the integrals are not absolutely convergent. Here the oscillatory nature of the amplitudes is crucial in order to obtain finite results.

\item We have thus also illustrated the usefulness of the Wynn algorithm in dealing with oscillatory integrals. The work~\cite{Dittrich:2023rcr} applied the Wynn algorithm successfully to sums arising from partition functions and expectation value calculations for effective spin foams~\cite{Asante:2020qpa,Asante:2021zzh}. It was pointed out in~\cite{Dittrich:2023rcr} that in order for the algorithm to work, it is important that the action scales at most linearly in the summation variables, which are the areas. We found here that the leading-order terms in the action are indeed linear in the areas. One should therefore be able to apply the Wynn algorithm in order to study $5-1$ and $4-2$ configurations in spin foams. 
\end{itemize}

These results provide valuable progress for our understanding of Lorentzian simplicial path integrals.
 In particular, the results on the finiteness show how Lorentzian Quantum Regge calculus can circumvent one of the crucial issues of Euclidean Quantum Regge calculus, namely the conformal mode problem. Here we do {\it not} need to employ a restriction on the signature of, e.g.~the edges, as proposed (for two-dimensional triangulations) in~\cite{Tate:2011ct}, or an exponentially suppressing measure as in~\cite{Mikovic:2023tgg}.

The finiteness of the partition function is also a key question for spin foams.  So-called bubble configurations, in which areas can become arbitrarily large, have been a major concern in spin foam research, e.g.~\cite{Perini:2008pd,Bonzom:2013ofa,Riello:2013bzw,Banburski:2014cwa,Chen:2016aag,Dona:2022vyh,Dona:2023myv} and for group field theories~\cite{BenGeloun:2011jnm,Finocchiaro:2020fhl,Carrozza:2024gnh}. For technical reasons, efforts have been focused on configurations which include only spacelike tetrahedra, but, as mentioned above, these do not correspond to Regge configurations with large areas. We have seen here that there are however many possible Regge configurations with unbounded bulk areas, if we allow timelike sub-simplices. 

We have found here, that the Regge action in the asymptotic regimes simplifies and in particular observed a certain type of dimensional reduction by one, and in some cases by two dimensions. A similar dimensional reduction has been found in spin foams~\cite{Riello:2013bzw} for a so-called melon configuration, where one has two 4-simplices glued along four tetrahedra. This work considered 4-simplices with only spacelike tetrahedra. It would be interesting to see whether this behaviour generalizes, and whether spin foam amplitudes simplify in the same way as we have found here for the Regge action.  

More generally, the work here motivates to consider systematically new asymptotic regimes for simplex amplitudes.
The resulting asymptotic expressions  might facilitate the numerical investigation of bubbles. Bubbles involve simplex amplitudes where some of the   areas are large and one has thus large spin numbers. Such amplitudes are however prohibitively computationally expensive~\cite{Dona:2022yyn}. Using possible simplifications in regimes where some areas are much larger than other areas, will allow a probe of a much richer regime of spin foams, without having to compute the full amplitudes explicitly. The idea of hybrid schemes, which so far have only been suggested for the regime where all areas are large~\cite{Asante:2022lnp}, could be extended to mixed regimes of large and small areas.

Another interesting application of dimensional reduction is the construction of $(d-1)$-dimensional quantum-geometric models from $d$-dimensional ones. A similar relationship has been observed for the three-dimensional Ponzano-Regge model and two-dimensional so-called intertwiner models~\cite{Dittrich:2013aia}, both of which are topologically invariant. Generalizing such relationships might also lead to new examples for holographic behaviour in quantum-geometric models~\cite{Bonzom:2015ans,Dittrich:2017hnl,Dittrich:2018xuk,Dittrich:2017rvb,Asante:2018kfo}.

As for three-dimensional Regge calculus, we have found also for the four-dimensional version a number of cases in which the asymptotic regime is light cone irregular. We can either exclude such configurations from the path integral (as done in Causal Dynamical Triangulations~\cite{Jordan:2013awa}) or include them, but then choose as integration contour the suppressing side of the branch cuts.  

Invariance of the quantum gravity partition function under changes of the triangulation is equivalent to an implementation of a discrete form of diffeomorphism invariance~\cite{Bahr:2011uj,Dittrich:2011ien}.  Therefore, a criterion for deciding which option to choose, is to look for (approximate) invariance of the partition function under $5-1$ and $4-2$ Pachner moves. This work provided valuable tools to do so.

\begin{acknowledgments}

The authors thank Jos\'e Padua-Arg\"uelles for providing  his derivation of the result in Section~\ref{GentoN}. JB is supported by an NSERC grant awarded to BD and a doctoral scholarship by the German Academic Scholarship Foundation. DQ is supported by research grants provided by the Blaumann Foundation.
Part of this research was conducted while BD was visiting the Okinawa Institute for Science and Technology (OIST) through the Theoretical Sciences Visiting Program (TSVP). Research at Perimeter Institute is supported in part by the Government of Canada through the Department of Innovation, Science and Economic Development Canada and by the Province of Ontario through the Ministry of Colleges and Universities.

\end{acknowledgments}

\bibliographystyle{jhep}
\bibliography{references}

\end{document}